\documentclass[fleqn,usenatbib]{mnras}

\usepackage{newtxtext,newtxmath}

\usepackage[T1]{fontenc}

\DeclareRobustCommand{\VAN}[3]{#2}
\let\VANthebibliography\thebibliography
\def\thebibliography{\DeclareRobustCommand{\VAN}[3]{##3}\VANthebibliography}

\usepackage{graphicx}	
\usepackage{amsmath, mathtools, nccmath, nicefrac}	
\usepackage{gensymb} 
\usepackage{enumitem}
\usepackage{textcomp}
\usepackage{chngpage}
\usepackage{footmisc}
\usepackage{array,longtable,pdflscape}
\usepackage[pro]{fontawesome5}
\usepackage{bm}
\usepackage{xspace}
\usepackage{tablefootnote, threeparttable}
\usepackage{float}
\usepackage{xcolor}
\usepackage{arydshln}

\definecolor{new_color}{HTML}{CF0000} 

\usepackage{natbib}

\PassOptionsToPackage{usenames, dvipsnames}{color}
\definecolor{dblue}{RGB}{40, 116, 166}
\definecolor{navy}{RGB}{56, 70, 184}
\definecolor{darkblue}{RGB}{39, 76, 119}
\definecolor{grey}{RGB}{255, 87, 51}
\definecolor{new_color}{HTML}{CF0000} 
\definecolor{crimson}{RGB}{220,20,60}

\newcommand\blue[1]{\textcolor{darkblue}{\textbf{#1}}}

\hypersetup{
    colorlinks = true,
    linkcolor = black,
    citecolor= dblue,
    urlcolor = cyan,
}

\newcommand{\Msun}{\mbox{$M_{\odot}$}}

\newcommand{\Rearth}{R$_{\oplus}$}

\newcommand{\Teff}{$T_{\rm eff}$}
\newcommand{\Teffmontreal}{$T_{\rm eff; MWDD}$}
\newcommand{\Teffmix}{$T_{\rm eff; mixed}$}
\newcommand{\logg}{$\log g$}

\newcommand{\upshapeI}[1]{\textrm{\upshape I}}
\newcommand{\Gaia}{\emph{Gaia}}
\newcommand{\angs}{\text{\normalfont\AA}}
\newcommand{\TESS}{\emph{TESS}}
\newcommand{\GALEX}{\emph{GALEX}}
\newcommand{\panstarrs}{\emph{Pan-STARRS}}
 
\newcommand{\logCaH}{$\log_{10}(\rm{Ca/H})$} 
\newcommand{\logCaHe}{$\log_{10}(\rm{Ca/He})$}

\title[Gaia EDR3/LAMOST Polluted White Dwarfs]{Detection and Preliminary Characterisation of Polluted White Dwarfs from \Gaia{} EDR3 and LAMOST}

\author[Badenas-Agusti et al.]{Mariona Badenas-Agusti$^{1,2,3, \thanks{\rm{E-mail: mbadenas@mit.edu}}}$,
Andrew Vanderburg,$^{2}$
Simon Blouin,$^{4}$
Patrick Dufour,$^{5}$
Javier Viaña,$^{2}$
\newauthor
Sara Seager,$^{1,2,6}$
Sharon X. Wang$^{7}$\\
$^{1}$Department of Earth, Atmospheric and Planetary Sciences, Massachusetts Institute of Technology, Cambridge, MA 02139, USA\\
$^{2}$Department of Physics and Kavli Institute for Astrophysics and Space Research, Massachusetts Institute of Technology, Cambridge, MA 02139, USA\\
$^{3}$MIT William Asbjornsen Albert Memorial Fellow\\
$^{4}$Department of Physics and Astronomy, University of Victoria, Victoria, BC V8W 2Y2, Canada\\
$^{5}$Département de Physique, Université de Montréal, Montréal, Québec H3C 3J7, Canada\\
$^{6}$Department of Aeronautics and Astronautics, Massachusetts Institute of Technology, Cambridge, MA 02139, USA\\
$^{7}${Department of Astronomy, Tsinghua University, Beijing 100084, China}\\ 
}

\date{Accepted 2023 October 29. Received 2023 October 28; in original form 2023 March 06}

\pubyear{2023}

\begin{document}
\label{firstpage}
\pagerange{\pageref{firstpage}--\pageref{lastpage}}
\maketitle

\begin{abstract}
We present a catalogue of 62 polluted white dwarfs observed by the 9th Low-Resolution Data Release of the Large Sky Area Multi-Object Fiber Spectroscopic Telescope (LAMOST LRS DR9v1; R$\approx$1,800) and the Early Data Release 3 (EDR3) of the \Gaia{} Mission. Among these stellar remnants, 30 are new discoveries with previously unknown traces of calcium pollution. To generate our catalogue, we used a database of 4,324 unique \Gaia{} EDR3 white dwarf candidates with LAMOST LRS DR9v1 observations, many of which have been spectroscopically confirmed by other telescopes. For these stars, we developed a quantitative method to detect calcium absorption in their spectra between 3,900-4,000~\angs, which we then validated through visual inspection and multiple literature cross-checks. Our catalogue provides the astrometric and photometric properties of the white dwarf candidates, incorporates supplementary data (e.g. Montreal White Dwarf Database, MWDD; PanSTARRS; the Hubble Space Telescope), and indicates the possibility of calcium pollution in their atmospheres. For our final sample of polluted white dwarfs, we also determine the main atmospheric properties of 23 sources with effective temperatures \Teff$\leq$25,000~K and no existing calcium abundances in the MWDD. Our analysis represents a first step towards measuring the full atmospheric composition of these stars and learning about the bulk properties of their accreted material. As we venture into the era of wide-field spectroscopic surveys, our work highlights the importance of combining large-scale databases for identifying and characterising new polluted white dwarfs.
\end{abstract}

\begin{keywords}
stars: white dwarfs, atmospheres, abundances -- astronomical data bases: catalogues -- techniques: spectroscopic -- methods: data analysis, observational
\end{keywords}



\section{Introduction}

The majority of stars in the Galaxy ($\sim$98$\%$), including our Sun, will end their days as white dwarfs \citep{Fontaine:2001}. These objects are the dense, degenerate remnants of low- and intermediate-mass Main-Sequence (MS) stars, that is, stars with initial masses lower than about $\leq$10~\Msun{} \citep{Weidemann:1983}. Having exhausted all their nuclear fuel, they are in a state of gradual cooling, slowly releasing the thermal radiation they accumulated over billions of years during their MS evolution. 

White dwarfs (WDs) are typically the size of the Earth ($\sim$1~\Rearth) with half the mass of the Sun ($\sim$0.6~\Msun). As a result, they have extremely high densities and surface gravities. Their strong gravitational fields cause their atmospheres to be chemically stratified: light elements (hydrogen and helium) are expected to remain in the thin outer layers of their atmospheres, while elements heavier than helium (or metals) should sink rapidly towards the core with diffusion timescales shorter than the evolutionary or cooling age of the white dwarf \citep{Althaus:2010}. In contrast to this expectation, however, spectroscopic observations of white dwarfs show that between 25$\%$ and 50$\%$ of them are ``polluted'' with heavy elements in their atmospheres \citep{Zuckerman:2003, Zuckerman:2010, Koester:2014}, particularly with calcium (Ca) and other rock-forming species such as magnesium (Mg), silicon (Si), and iron (Fe). Given the short diffusion timescales of metals \citep{Paquette:1986, Koester:2009}, {their presence in the atmospheres of some white dwarfs suggests that they are not primordial, but exogeneous --in other words, that they must have recently been accreted by the star.\footnote{Some low- and high-temperature white dwarfs may also show spectral features from other physical mechanisms. For white dwarfs with effective temperatures higher than \Teff$\gtrapprox$25,000~K, the presence of heavy elements may be caused by the outward effect of radiative levitation pressure \citep{Chayer:1995}. In cool, hydrogen-deficient white dwarfs, carbon pollution may be due to convective dredge-up from the deep stellar interior \citep{Pelletier:1986, Camisassa:2017, Bedard:2022b}.} 

To date, the standard theory of {most white dwarf metal pollution is the recent or ongoing accretion of asteroid- or comet-like planetesimals that survived the post-MS evolution of their host star \citep[e.g., see the review by][]{Jura:2014}. Growing observational evidence supports this canonical interpretation, including the detection of X-rays from ongoing accretion \citep{Cunningham:2022}, near infrared excesses from dust disks \citep{Zuckerman:1987, Jura:2003}, double-peaked optical emission features from gas disks \citep{Gansicke:2006}, and heavy element absorption lines imprinting the composition of disintegrating planets \citep{Zuckerman:2007} and close-in planetesimals \citep{Vanderburg:2015, Manser:2019}.

\begin{figure*}
   \centering
   \includegraphics[width=1\textwidth]{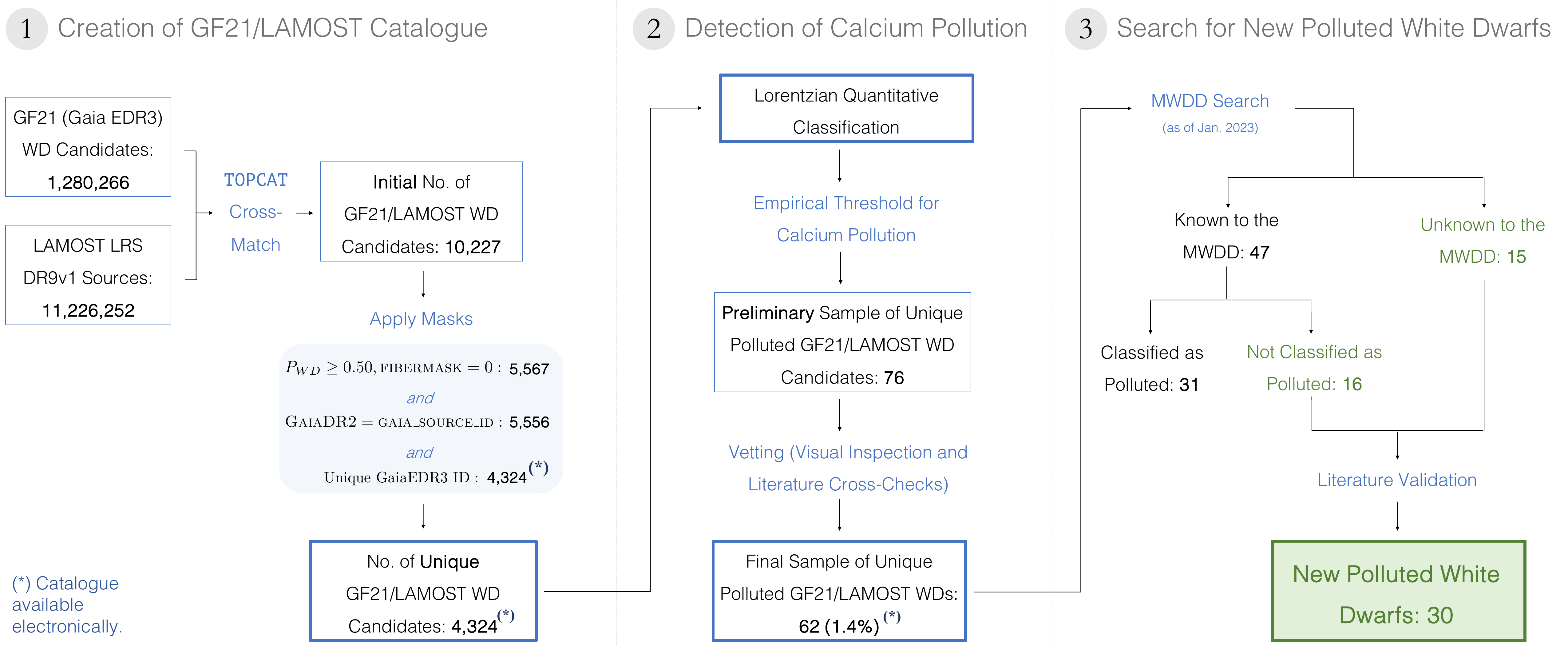}
   \caption{A summary of our methodology for generating a catalogue of \Gaia{} EDR3 white dwarf candidates with LAMOST low-resolution spectra, and identifying those with calcium pollution between 3,900-4,000~\angs. First, we cross-matched the \Gaia{} EDR3 database of \citealt{GentileFusillo:2021} (GF21) with the LAMOST LRS DR9v1 archive, obtaining an initial sample of 10,227 stars. We then restricted the latter to 4,324 unique white dwarf candidates by imposing (i) a GF21 probability greater than $P_{\rm WD}\geq0.50$, (ii) no fiber issues in the LAMOST spectrograph (\textsc{fibermask=0}), (iii) consistent \Gaia{} identifiers (\textsc{GaiaDR2}=\textsc{gaia\_source\_id}), and (iv) a unique LAMOST match per GF21 star (Section \ref{sec:methodology_creation_catalog}). Next, we used a quantitative model to create a preliminary sample of 76 polluted white dwarf candidates (Section  \ref{sec:lorentz_classification} and Section \ref{sec:pollution_threshold}). After thoroughly vetting this sample through both visual and literature cross-checks (Section \ref{sec:literature_validation}), we generated a final catalogue of 62 polluted white dwarfs (1.43$\%$), of which 30 represent new discoveries at the time of this writing (Section \ref{sec:final_polluted_sample}).}
   \label{fig:flowchart}
\end{figure*}

Since the discovery of 17 heavy elements in the atmosphere of the white dwarf GD 362 \citep{Zuckerman:2007}, high-resolution spectroscopy of polluted white dwarfs has become the most direct and accurate way of constraining the bulk composition of their accreted material \citep{Jura:2014, Veras:2021}. As of 2023, this technique has enabled the detection of more than 20 different metals \citep{Klein:2021}, revealing that the majority of observed pollutants are volatile-poor and rocky \citep{Klein:2010, Xu:2019, Doyle:2020}, with a few water-rich exceptions \citep[e.g.][]{Farihi:2011, Raddi:2015, Xu:2017}. Despite these important discoveries, only a few dozen out of more than 1,000 known polluted white dwarfs have been studied in detail with high-resolution spectroscopy \citep[e.g.][]{Xu:2019, Putirka:2021}. 

The scarcity of polluted white dwarfs with well-measured photospheric abundances has hindered our understanding of their polluting bodies and of rocky exoplanetary material in general. To address this limitation, we can combine the large databases of wide-field astronomical surveys to identify the best targets for high-resolution spectroscopic work. In this paper, we intend to contribute to this goal by presenting a new catalogue of polluted white dwarfs observed by both the Early Data Release 3 (EDR3) of the \Gaia{} Mission and the 9th Low-Resolution Data Release (LRS DR9v1) of the Large Sky Area Multi-Object Fiber Spectroscopic Telescope (LAMOST; \citealt{Zhao:2012, Cui:2012}).\footnote{\url{http://www.lamost.org/public/}}

As of today, LAMOST constitutes the largest low-resolution spectroscopic survey in operation. However, despite containing millions of stellar spectra, it has remained largely unexplored for the study of polluted white dwarfs. With our work, we aim to reverse this trend by exploiting the rich dataset of LAMOST to: (i) build the first catalogue of \Gaia{} EDR3 white dwarf candidates with existing low-resolution LAMOST spectroscopy, (ii) identify those sources with traces of calcium pollution; and (iii) determine the main physical properties and chemical abundances of a subsample of polluted white dwarfs with effective temperatures \Teff$\leq$25,000~K and no existing calcium abundances in the Montreal White Dwarf Database (MWDD; \citealt{Dufour:2016_MWDD}). For this atmospheric analysis, we imposed a temperature cut at \Teff$\leq$25,000~K to ensure that the presence of metal pollutants was due to accretion of exoplanetary material, rather than to radiative levitation.

This paper is organised as follows. In Section \ref{sec:data}, we describe the observations used in this work. Section \ref{sec:methodology} explains our framework for building a catalogue of \Gaia{} EDR3 white dwarf candidates with LAMOST LRS DR9v1 spectra (\autoref{fig:flowchart}). In Section \ref{sec:analysis}, we summarise the main properties of these stars, identify those with calcium pollution, and present the results of our atmospheric analysis. Finally, we discuss and conclude in Section \ref{sec:discussion_and_conclusions}.

\section{Observational Data} \label{sec:data}

\subsection{The LAMOST Spectroscopic Survey}

The Large Sky Area Multi-Object Fibre Spectroscopic Telescope -- also known as the Guo Shou Jing Telescope -- is located in Xinglong (China) and is maintained by the National Chinese Academy of Sciences. This instrument is a reflecting Schmidt telescope with a variable effective aperture of 3.6 to 4.9 meters (m), a focal length of 20~m, a wide field-of-view of 5$\degree$, and two spectroscopic modes (low and medium) with a resolving power ($R\equiv\lambda/\Delta\lambda$) of R$\approx$1,800 and R$\approx$7,500, respectively. The wavelength coverage and magnitude range of each mode is listed in Table \ref{tab:LRS_vs_MRS}.

\begin{table}
    \centering
    \caption{Resolving power, wavelength coverage, and magnitude range of LAMOST' low- and high-resolution spectroscopic modes \citep{Zhao:2012, Yan:2022}. \label{tab:LRS_vs_MRS}}
    \begin{tabular}{lcc}
        \hline
        \noalign{\smallskip}
        Property & Low (LRS)                & Medium (MRS)  \\ 
         \hline
        \hline
        Resolving Power ($R$)   & 1,800  & 7,500 \\ 
        Coverage (\angs)   & [3,690-9,100]  &  [4,950-5,350], [6,300-6,800] \\ 
        Magnitude          & [9, 19]~$r_{\rm mag}$ &  [9, 15]~$G_{\rm mag}$ \\
        \hline
    \end{tabular}
\end{table}

Since the start of its pilot survey over a decade ago, LAMOST has scanned the northern hemisphere sky, acquiring more than 17 million spectra of stars, galaxies, and other astrophysical sources, including quasars and black holes \citep{Yan:2022}. With these observations, LAMOST has become one of the largest multi-object spectroscopic surveys to date, hence playing an important role in furthering our understanding of the Universe. The high efficiency and observing cadence of LAMOST is mainly due to its complex instrumentation system: with 16 identical spectrographs, 32 Charge-Coupled Devices (CCDs), and 4,000 circular fibers of 3.3 arcseconds (\arcsec) in diameter mounted on its focal plane, LAMOST can observe up to 4,000 stars in a single exposure with a limiting magnitude of $r_{\rm mag}$=19 at R$\approx$1,800 \citep{Zhao:2012}. 

As of this writing, the LAMOST database\footnote{\url{http://dr.lamost.org/data-release.html}} consists of 10 \textit{cumulative} Data Releases or ``DR'' (see Table \ref{tab:lamost_dr}), of which eight are public (DR1-DR8), and two are internal (DR9-DR10). These DRs contain pre-processed data produced by the LAMOST analysis pipeline, also known as ``1D Pipeline'' \citep{Luo:2012}. For our work, we downloaded the low-resolution spectra from DR9v1 as of August 2022.

\subsection{The \Gaia{} EDR3 Mission} 

The \Gaia{} Mission of the European Space Agency was designed to produce the largest and most accurate three-dimensional map of the Milky Way \citep{Gaia:2016}. Since its launch in 2013, \Gaia{} has been scanning the sky to achieve this goal, providing important observations for the study of stellar and planetary astrophysics. In this paper, we use the catalogue of white dwarf candidates from \citealt{GentileFusillo:2021} (``GF21'' hereafter), which relies on \Gaia{} EDR3 data.

\begin{table}
    \centering
    \caption{Properties of LAMOST Data Releases.$^{\star}$ The number of available stellar spectra was last updated in January 2023.}
    \label{tab:lamost_dr}
    \begin{tabular}{lccc}
        \hline
        DR$^{\star}$ & Version  & Stellar Spectra & Public/Internal   \\ 
        \hline  \hline
	\multicolumn{4}{c}{\it Phase 1 Survey (2012-2017)}  \\
        \hline
	DR1        & -  & 1,061,918  & Public \\
        DR2        & -  & 2,207,189  & Public \\
        DR3        & -  & 3,177,995  & Public \\
        DR4        & v2 & 4,537,436  & Public \\
        DR5        & v3 & 5,348,712  & Public \\
        \hline
        \multicolumn{4}{c}{\it Phase 2 Survey (2017-)}\\
        \hline 
               & v2 & 9,911,337  & Public \\
        DR7    & v2 & 10,431,197 & Public \\
        DR8    & v2 & 10,633,515 & Public \\     
        DR9    & v1 & 11,226,252 & Internal \\
        DR10   & v0 (Q1-Q3)   & 87,632     & Internal \\
        \hline
    \end{tabular}
    \begin{quote}
        \hspace{0.1pt} $^{\star}$ \url{http://dr.lamost.org/data-release.html}
    \end{quote}
\end{table}
Released in December 2020, \Gaia{} EDR3 contains accurate astrometric and photometric measurements for about 1.8 billion point sources with $G_{\rm mag}$-magnitudes in the range $3\leq G_{\rm mag}\leq21$, including their positions (Right Ascension ``R. A.'' and Declination ``Dec'' at epoch J2016.0), proper motions, and parallaxes \citep{Gaia:2021_edr3}. In contrast to \Gaia{}'s Data Release 2 (DR2), \Gaia{} EDR3 uses a longer observational baseline (34 months vs. 22 months for DR2), its parallax uncertainties are 20-30$\%$ smaller, and its proper motion measurements are twice as precise \citep{Lindegren:2021, Riello:2021}.

\begin{figure*}
   \centering
   \includegraphics[width=1\textwidth]{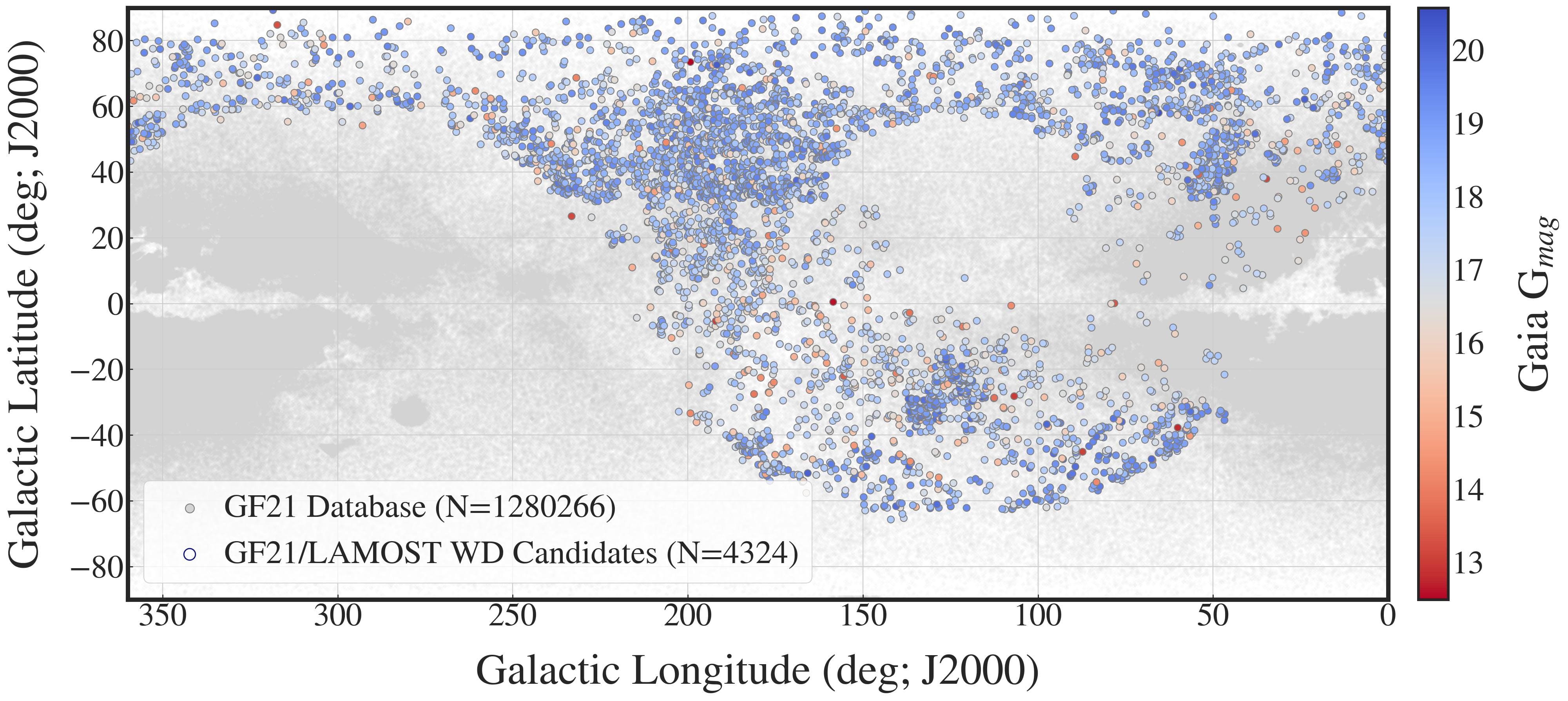}
   \caption{Galactic coordinates (latitude and longitude in J2000) of 4,324 \Gaia{} EDR3 white dwarf candidates with existing LAMOST LRS DR9v1 spectra, with their colour representing their \Gaia{} G-magnitude. The sample of white dwarf candidates from \citealt{GentileFusillo:2021} is also shown in light grey. }
   \label{fig:sky}
\end{figure*} 

In GF21, the authors used the \Gaia{} EDR3 database to create a catalogue of 1,280,266 sources in the WD region of the Hertzsprung-Russell (H-R) diagram. Their catalogue contains all the columns from the \Gaia{} EDR3 archive as well as additional columns with parameters derived by GF21. Among the latter is the probability that an object is a white dwarf ($P_{\rm WD}$), which depends on the location of the star in the H-R diagram. According to GF21, $P_{\rm WD}\geq0.75$ represents a general threshold for high-confidence white dwarfs with a minimal level of contamination. Based on this criterion, GF21 identified 359,073 white dwarf candidates, offering the largest public collection of such objects to date \citep{GentileFusillo:2021}.

\subsection{The Montreal White Dwarf Database}

The Montreal White Dwarf Database\footnote{\url{https://www.montrealwhitedwarfdatabase.org/}} \citep{Dufour:2016_MWDD} contains most of the spectroscopically confirmed white dwarfs known to date. For our paper, we used the MWDD as of January 2023, which features 68,374 white dwarfs described in more than 150 published manuscripts. 

\begin{figure*}
    \centering
    \includegraphics[width=0.80\linewidth]{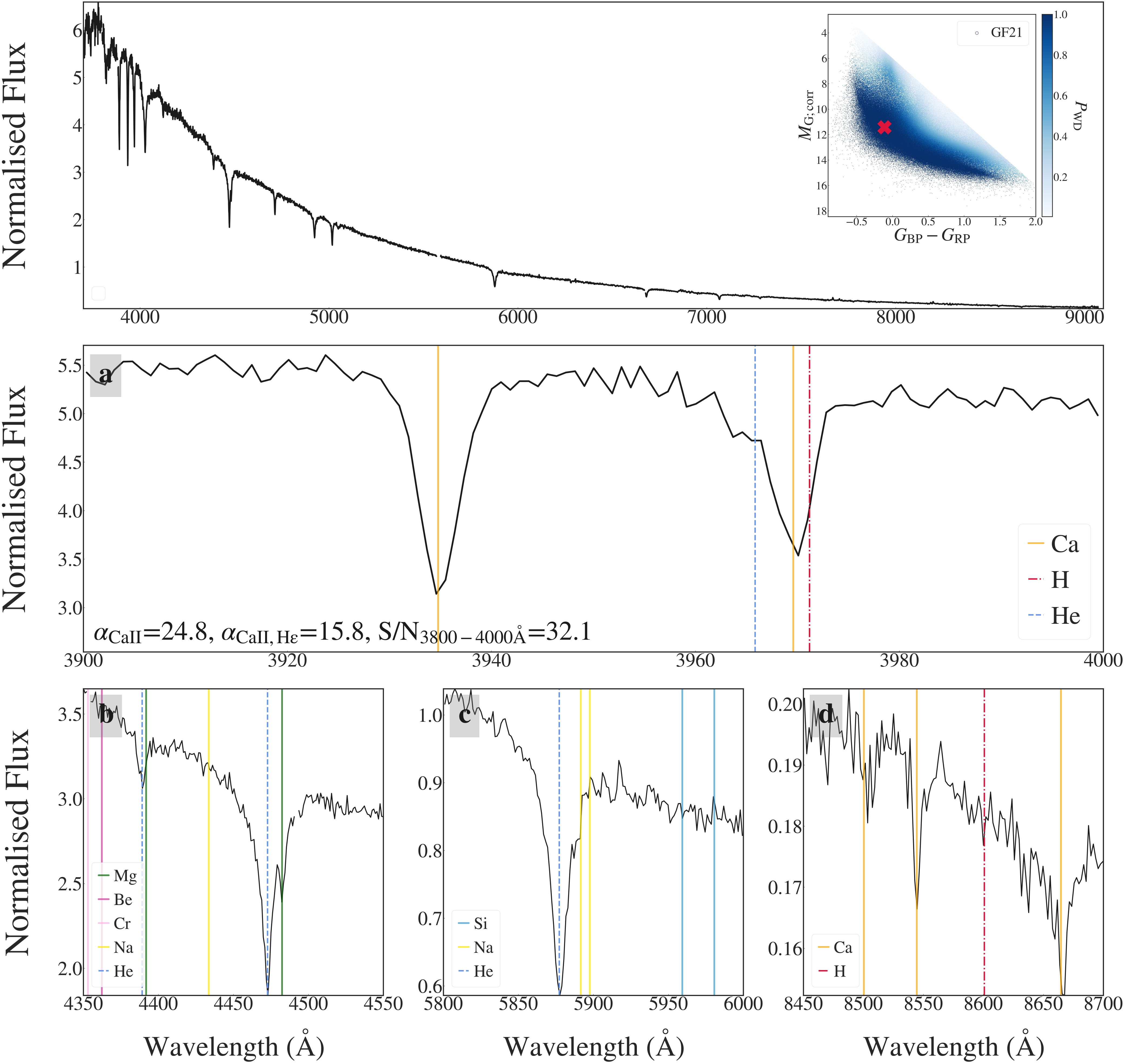}
    \caption{\textit{Top panel}: LAMOST low-resolution spectrum of the He-rich polluted white dwarf \Gaia{} EDR3 5187830356195791488, also known as GD 40 \citep{Klein:2010}. The red cross in the inset plot shows its absolute corrected \Gaia{} magnitude and its G$_{BP}$-G$_{RP}$ colour in an H-R diagram with the GF21 stars. \textit{Panels (a)-(d):} Zoomed-in regions of the spectrum covering multiple metallic lines. In panel \textit{(a)}, we also show the S/N of the region between 3,800~\angs{} and 4,000~\angs{}, as well as the significances of the Ca II and blended Ca II/H$\epsilon$ lines at about 3,934~\angs{} and 3,969.5~\angs, respectively (see Section \ref{sec:lorentz_classification}).}
   \label{fig:gd40}
\end{figure*}

\section{Methodology} \label{sec:methodology}

\subsection{Creation of a White Dwarf Catalogue}\label{sec:methodology_creation_catalog}

The first step in our analysis was to search for GF21 white dwarfs candidates with existing LAMOST spectroscopy. To this end, we started by downloading the {GF21 database\footnote{\url{https://vizier.cds.unistra.fr/viz-bin/VizieR?-source=J/MNRAS/508/3877} and the LAMOST low-resolution DR9v1\footnote{\url{http://www.lamost.org/dr9/v1.0/}} Input Catalogue.}\footnote{The LAMOST Data Releases are composed of a ``General Catalogue'' and/or an ``Input Catalogue.'' The ``Input Catalogue'' contains all the objects that LAMOST was originally commissioned to observe, with only a fraction of them eventually observed by the spectrograph.} Next, we used the \texttt{TOPCAT}\footnote{\url{http://www.star.bris.ac.uk/~mbt/topcat/}.} software \citep{Taylor_2017} to cross-match the R.A. and Dec. of the GF21 and LAMOST sources, taking J2000 as our reference equinox and correcting the positions of the stars to epoch J2000 using their proper motions. During our cross-match routine, we allowed multiple LAMOST matches per GF21 star and assumed a maximum sky separation of 1.5\arcsec{} between two objects, which corresponds to about half of the diameter of the LAMOST circular fibers (3.3\arcsec). Under these assumptions, we identified a total of 10,227 \Gaia{} EDR3 white dwarf candidates with LAMOST low-resolution observations.

Taking advantage of multiple GF21 and LAMOST quality flags, we then imposed three conditions on our catalogue: (i) $P_{\rm WD}\geq0.50$, which allowed us to discard low-probability white dwarfs; (ii) \textsc{fibermask}=0, a LAMOST mask indicating the correct operation of the spectrograph's fibers; and (iii) \textsc{Gaia DR2=gaia\_source\_id}, where ``\textsc{Gaia DR2}'' is the \Gaia{} DR2 identifier in the GF21 database, and ``\textsc{gaia\_source\_id}'' is the equivalent parameter for LAMOST. These three criteria reduced our catalogue to 5,556 white dwarf candidates, of which 5,092 had been previously assigned the ``\texttt{WD}'' label by the LAMOST 1D Pipeline. According to this pipeline, the remaining 464 sources included a variety of astrophysical objects, mostly hot A/F-type stars (see \ref{tab:non_wds} in the Appendix). To avoid losing any potential white dwarfs, we opted for including them all in our catalogue.  

Given the cumulative nature of the LAMOST survey, our initial catalogue of 5,556 stars contained multiple repeated entries for those GF21 targets that had been observed more than once by LAMOST. For this paper, we chose to restrict our database further by only considering a single LAMOST spectrum per GF21 star. To identify the most suitable spectrum for our analysis, we proceeded as follows: first, we identified those GF21 sources with multiple LAMOST observations; then, for each spectrum, we computed its mean signal-to-noise (S/N) by averaging  its S/N metrics in the LAMOST database (see columns \textsc{snru}, \textsc{snrg}, \textsc{snrr}, \textsc{snri}, \textsc{snrz} in Table \ref{tab:catalog_format}); finally, we only kept the LAMOST spectrum with the highest mean S/N and removed the remaining observations from our catalogue.\footnote{We also contemplated the possibility of co-adding all the existing LAMOST spectra for a given white dwarf. Nevertheless, we found that when we included the lowest S/N data, there were often instrumental artefacts that degraded the quality of the final co-added spectra. Therefore, our catalogue uses the highest S/N spectrum instead.} To ensure the traceability of the discarded spectra, however, we classified the latter with two parameters from our \texttt{TOPCAT} cross-match routine, namely: the ``\textsc{GroupSize}'' (indicating the number of LAMOST matches per GF21 star), and the ``\textsc{GroupID}'' (an identification number shared by all the LAMOST matches associated to a specific GF21 star). Both parameters are included in our catalogue for reference purposes. 

The requirement of \textit{unique} matches between GF21 and LAMOST yielded a final GF21/LAMOST database of 4,324 white dwarf candidates with $P_{\rm WD}\geq0.50$, of which 891 (20.6$\%$) are not included in the MWDD as of January 2023. The spatial distribution of these stars is illustrated in \autoref{fig:sky}, together with the galactic coordinates of the GF21 sources. Given that \Gaia's angular resolution ($\sim$0.4\arcsec) is significantly finer than that of the LAMOST circular fibers ($\sim$1.5\arcsec), we concluded this section by exploring the possibility of light contamination from nearby \Gaia{} objects. To this end, we used the CDS XMatch tool\footnote{\url{http://cdsxmatch.u-strasbg.fr/}.} to cross-match the coordinates of the 4,324 GF21/LAMOST white dwarf candidates with those of the \textit{entire} \Gaia{} EDR3 archive, using a maximum sky separation of 1.5\arcsec. From this analysis, we identified 41 GF21/LAMOST stars with \Gaia{} companions falling on the same LAMOST spectroscopic fiber (see ``\textsc{contamination\_1.5arcsec}'' column in our catalogue). Among them, only one is part of our final sample of polluted white dwarfs obtained in Section \ref{sec:final_polluted_sample} (\Gaia{} EDR3 203931163247581184, with an object 1.24\arcsec{} away). 

\subsection{Acquisition of LAMOST Spectra} 

After generating our catalogue of 4,324 GF21/LAMOST white dwarf candidates, we loaded their unique LAMOST \textsc{obsid} identifiers into the LAMOST DR9v1 search platform and retrieved their spectra as FITS files.

In the LAMOST DR9v1 database, FITS files share three important characteristics \citep{Luo:2015}. First, they are stored with the name ``\textit{spec-MMMMM-YYYY$\_$spXX-FFF.fits},'' where \textit{MMMMM} is the local modified Julian day, \textit{YYYY} is the plan identity string, \textit{XX} is a number between 1 and 16 indicating the spectrograph ID, and \textit{FFF} is a number between 1 and 250 representing the  LAMOST fiber number. These parameters can be found, respectively, in the ``\textsc{lmjd}'', ``\textsc{planid}'', ``\textsc{spid}'', and ``\textsc{fiberid}'' columns of our catalogue. Second, their headers provide useful keywords ranging from observation and weather data (e.g. start and end time of the observation, name and magnitude of the object, temperature of the CCD camera), to data reduction and spectral analysis parameters (e.g. extraction method of the observations, number of valid exposures, redshift of the object).\footnote{See \citealt{Luo:2015} for a full list of keywords.} Lastly, LAMOST FITS files contain six scientific products:\footnote{We recommend caution when extracting LAMOST scientific products, as they can appear in different row numbers in other Data Releases.} wavelength, raw flux, continuum-normalised flux, inverse variance of the flux uncertainties (i.e. one over sigma-squared),\footnote{Therefore, the signal-to-noise ratio of a pixel can be estimated with flux$\times$(inverse variance)$^{0.5}$} and two masks to evaluate the quality of each pixel (\textsc{Andmask} and \textsc{Ormask}).

\begin{figure*} 
    \begin{center}
        \includegraphics[width=0.82\textwidth]{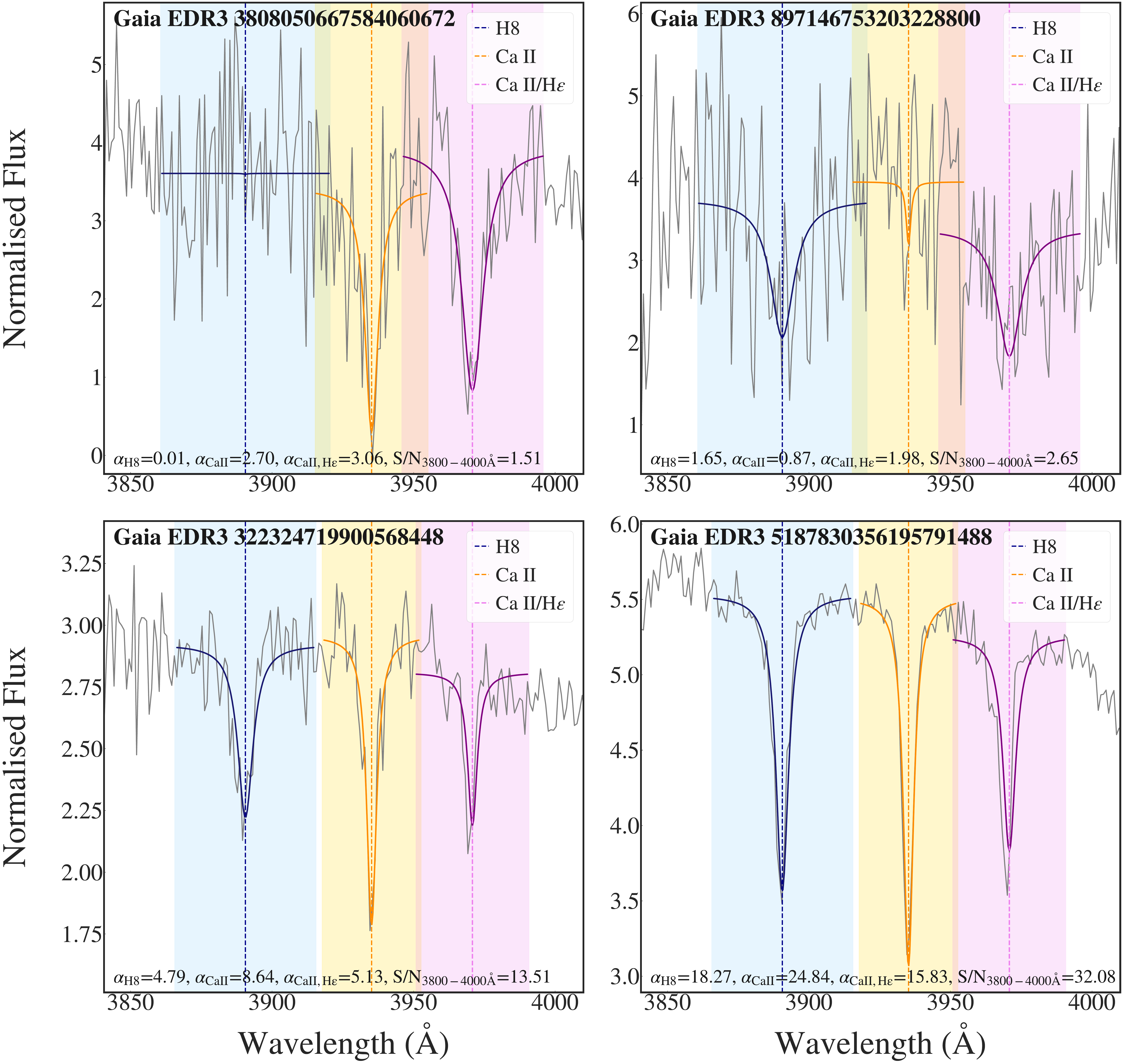}
        \caption{Lorentzian models and statistical significances (``$\alpha$'') of the H8 ($\sim$3,980~\angs), Ca II ($\sim$3,934~\angs) and blended Ca II/H$\epsilon$ absorption lines ($\sim$3,969.5~\angs) for 4 white dwarfs with different amounts of calcium pollution. At the bottom of each panel, we also include the S/N of their LAMOST spectra in the wavelength region between 3,800~\angs{} and 4,000~\angs{}.}
        \label{fig:lorentzian_fit}
    \end{center}
\end{figure*}

After downloading the LAMOST FITS files, we developed an automated \texttt{Python} routine to extract their corresponding scientific products and produce their low-resolution spectra. To improve the quality of the latter, we removed outlier points with the LAMOST \textsc{Andmask} and \textsc{Ormask} quality masks, and set to \textit{NaN} both infinite and negative flux values, as well as those points located in the échelle overlap region of LAMOST (5,700-5,900~\angs). \autoref{fig:gd40} shows the processed spectrum of \Gaia{} EDR3 5187830356195791488, a helium-dominated white dwarf also known as GD 40 \citep{Klein:2010}. The depth of the observed calcium lines between 3,900~\angs{} and 4,000~\angs{} reveals the heavily polluted nature of this object, while also demonstrating the ability of LAMOST to detect photospheric pollution at low resolution. Another feature of \autoref{fig:gd40} is the small deviation of the metallic lines relative to their expected central wavelengths. This phenomenon, which arises from the  barycentric radial velocity and gravitational redshift of the star, is accounted for in our spectral fitting routine (see Section \ref{sec:atmospheric_analysis}) to ensure that it does not bias our atmospheric models and retrieved calcium abundances.

\begin{figure*}
    \begin{center}
    \includegraphics[width=0.84\textwidth]{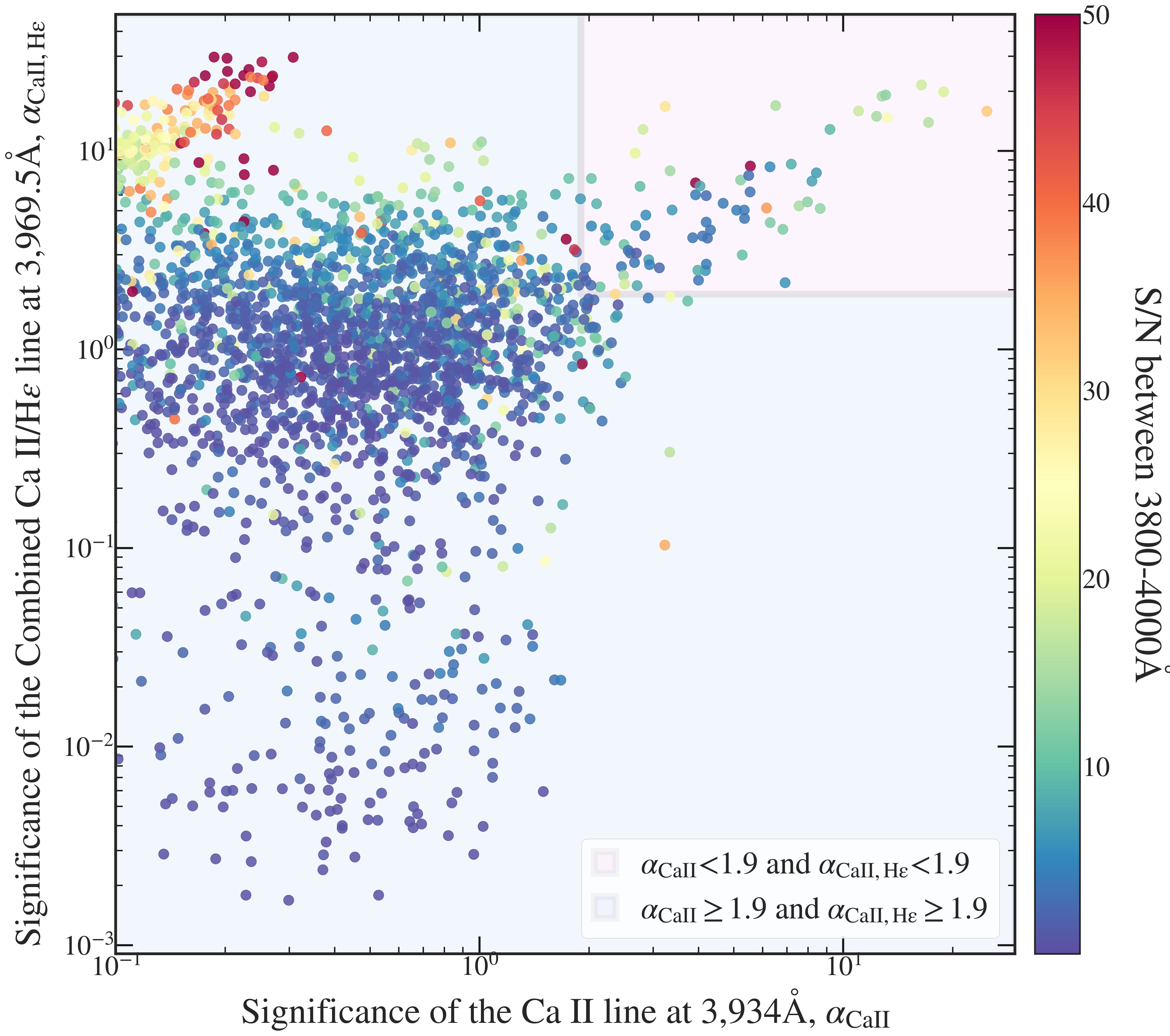}
    \caption{Statistical significances ($\alpha$) of the Ca II and blended Ca II/H$\epsilon$ lines at about 3,934~\angs{} and 3,969.5~\angs{}, respectively. Assuming a quantitative pollution threshold of $\alpha\geq1.90$ for both calcium lines (see shaded red region), we identify 76 white dwarfs with calcium signatures in their LAMOST spectra. In this figure, we only show our Lorentzian results for $\alpha\geq0.1$, with lower significances corresponding to observations below the noise level and with no detections. The colorbar illustrates the S/N of the LAMOST spectra between 3,800~\angs{} and 4,000~\angs{}, with red circles representing stars with a true S/N greater than 50.}
    \label{fig:significances}
    \end{center}
\end{figure*}

\subsection{Main Catalogue Properties}\label{sec:properties_catalogue}

Our GF21/LAMOST catalogue is characterised by 228 columns obtained from four different sources: the LAMOST DR9v1 archive, the \Gaia{} EDR3 database of GF21, the MWDD, and this work. The full catalogue can be downloaded in machine-readable format from our GitHub repository,\footnote{\url{https://github.com/mbadenas/gaialamost}} with Table \ref{tab:catalog_example} providing a selection of columns to illustrate its general format.

In addition to offering valuable information from \Gaia{} EDR3, LAMOST, and other external databases, our catalogue contains a number of additional metrics for each white dwarf, which we summarise in Table \ref{tab:catalog_format} in the Appendix. Among them, we report:

\begin{itemize}[noitemsep,topsep=1pt,leftmargin=14pt]
    \item information from our \texttt{TOPCAT} cross-match routine, such as the star's \textsc{GroupSize} and \textsc{GroupID} parameters, or the sky separation between its \Gaia{} EDR3 and LAMOST coordinates (see ``\textsc{GaiaLAMOST\_Separation}'' column), 
    \item the S/N of its LAMOST spectrum between 3,800-4,000~\angs{} as calculated in Section \ref{sec:lorentz_classification} (see column ``\textsc{snr\_ca\_window}''), as well as the mean S/N of its full spectrum, estimated by averaging all the S/N measurements in the LAMOST \textsc{u}, \textsc{g}, \textsc{r}, \textsc{i}, and  \textsc{z} bands (see column ``\textsc{lamost\_snr\_mean}''),
    \item its angular separation relative to the nearest star in the latest Data Releases of the Galaxy Evolution Explorer Mission (i.e. \GALEX{} GR6+7, \citealt{Martin:2005}), the Panoramic Survey Telescope $\&$ Rapid Response System imaging survey (\panstarrs{} DR2; \citealt{PanSTARRS:2010}), and the Input Catalogue (TIC) of the Transiting Exoplanet Survey Satellite (\TESS{} TICv8; \citealt{Ricker:2014}),
    \item its magnitudes and uncertainties in the TIC, \GALEX, and \panstarrs{} databases,
    \item the number of existing observations acquired by the Hubble Space Telescope (HST) with the STIS/CCD instrument, and
    \item a quantitative and visual assessment of the presence of calcium pollution in its LAMOST spectrum, following the conclusions derived from Section \ref{sec:analysis}.
\end{itemize}

For each white dwarf candidate, we downloaded its photometric magnitudes and number of existing HST observations with \texttt{Python}'s \textit{astroquery} package, which provides a direct interface with the Mikulski Archive for Space Telescopes (MAST).\footnote{\url{https://science.nasa.gov/astrophysics/astrophysics-data-centers/multimission-archive-at-stsci-mast}} We also used the default search radius of \textit{astroquery} to find its nearest star in the databases of \TESS, \panstarrs, and \Gaia{} (i.e. 0.02 degrees for \TESS{} and \panstarrs, and 0.05 arcminutes for \GALEX).

\section{Analysis} \label{sec:analysis} 

\subsection{Quantitative Assessment of Calcium Pollution} \label{sec:lorentz_classification}

At low-to-intermediate resolution, the majority of known polluted white dwarfs exhibit one or two heavy elements in their spectra, with calcium being one of the most detectable chemical species in the optical \citep{Coutu:2019}. Therefore, recognising the importance of calcium as a proxy for white dwarf pollution, we developed a quantitative method to detect calcium signatures in LAMOST low-resolution observations.

For this quantitative analysis, we employed Lorentzian functions to fit three important absorption lines between 3,800~\angs{} and 4,000~\angs{}, namely: hydrogen (H8) at an air wavelength of about 3,889.1~\angs, calcium (Ca II) at 3933.7~\angs{}, and the blended line of calcium (Ca II) and hydrogen (H$\epsilon$) at an average air wavelength of $\sim$3,969.5~\angs{} \citep{NIST_ASD}. To implement our algorithm, we started by median-normalising the LAMOST spectra and applying a 30~km/s shift to the center of the lines, which corresponds to a standard gravitational redshift for a 0.6~\Msun, 1~\Rearth{} white dwarf \citep{Chandra:2020}. We then optimised the Lorentzian function, given by: 

\begin{equation}
\label{eq:lorentzian}
f(\lambda) = a + \frac{b}{1 + \left(\frac{\lambda - c}{d}\right)^2},
\end{equation}
where $a$, $b$, and $d$ are the vertical offset, depth, and width of the lines, and $c$ is the location of the line center. We fitted for $a$, $b$, and $d$, while treating the shifted line center ($c$) as a fixed parameter. Although we also considered fitting the latter, the typical barycentric radial velocity and gravitational redshift of a white dwarf would only lead to small wavelength deviations (of the order of 1-2~\angs{}) compared to the resolution of LAMOST ($\approx$ 2.2~\angs{} in the calcium region between 3,800~\angs{} and 4,000~\angs{}). We thus chose to keep the shifted line centers fixed, obtaining better results by not letting them move freely away from their expected central wavelengths.  

After concluding our fitting routine, we estimated the statistical significances (``$\alpha$'' thereafter) of the H8, Ca II, and Ca II/H$\epsilon$ lines by calculating the ratio between their best-fit Lorentzian depth and the median of their median-normalised flux error in a window of 50~\angs, 35~\angs, and 40~\angs, respectively, around their central wavelength (see columns ``\textsc{sig\_H}'', ``\textsc{sig\_Ca}'', ``\textsc{sig\_CaH}'' in our catalogue). \autoref{fig:lorentzian_fit} illustrates our results for several polluted white dwarfs with varying levels of calcium pollution. In \autoref{fig:significances}, we also present the significances of the Ca II and combined Ca II/H$\epsilon$ lines for the white dwarf candidates in our catalogue, with the colorbar indicating the S/N of their spectra between 3,800~\angs{} and 4,000~\angs{} (see column ``\textsc{snr\_ca\_window}'' in Table \ref{tab:catalog_format}).

\subsection{Empirical Threshold of Calcium Pollution} \label{sec:pollution_threshold}

After estimating the calcium line significances for all the white dwarf candidates in our catalogue, we visually inspected their LAMOST spectra to better understand the general patterns observed in \autoref{fig:significances}, the sensitivity of our Lorentzian algorithm to calcium pollution, and its vulnerability to spurious detections. From this visual inspection, we found that the majority of observations were too noisy to confirm or rule out the presence of metal pollutants. However, guided by our visual inspection, and using the significance results from our Lorentzian analysis (see \autoref{fig:significances}), we established an approximate pollution threshold at $\alpha\geq1.90$ for the significances of the Ca II ($\alpha_{\mathrm{CaII}}$) and combined Ca II/H$\epsilon$ lines ($\alpha_{\mathrm{CaII,H\epsilon}}$), with the latter serving as tool to minimise our false positive detections.

Although the detection of metal pollution constitutes a multi-faceted problem and our cut at 1.90 may not include the most weakly calcium-polluted sources in our catalogue, we found this threshold to be the lowest at which we could detect calcium lines without contaminating our polluted sample with non-detections and noisy spectra. Therefore, under the assumption that a LAMOST spectrum is calcium-polluted if both $\alpha_{\mathrm{CaII}}\geq1.90$ and $\alpha_{\mathrm{CaII,H\epsilon}}\geq1.90$ are satisfied, we identified a total of 76 white dwarf candidates with traces of calcium at about 3,934~\angs{} and 3,969.5~\angs{} (see shaded red area in \autoref{fig:significances}).

\subsection{Literature Validation} \label{sec:literature_validation}

To validate and confirm our preliminary sample of polluted white dwarf candidates, we carefully inspected their LAMOST spectra and performed extensive literature cross-checks with the MWDD, Simbad's Astronomical Database,\footnote{\url{http://simbad.cds.unistra.fr/simbad/}} and existing spectroscopic observations from other astronomical surveys. In this section, we describe our vetting steps for generating a final sample of polluted white dwarfs. 

\subsubsection{Visual Inspection of LAMOST Spectra} \label{sec:visual_check}

First, we visually examined the LAMOST spectra of the 76 polluted stars, identifying three objects with spectral features atypical of isolated white dwarfs (\autoref{fig:discard_visual}), namely: \Gaia{} EDR3 1289860214647954816 ($P_{\rm{WD}}=0.53$, \citealt{Drake:2014}), \Gaia{} EDR3 880821067114616832 ($P_{\rm{WD}}=0.61$, \citealt{Denisenko:2012}), and \Gaia{} EDR3 710040763560788608 ($P_{\rm{WD}}=0.67$, \citealt{Gaia:2018}). The LAMOST spectra of both \Gaia{} EDR3 1289860214647954816 and \Gaia{} EDR3 880821067114616832 exhibit multiple emission lines between 3,600~\angs{} and 7,000~\angs{}, which suggests that they are not single white dwarfs, but rather cataclysmic variables as proposed by \citealt{Drake:2014} and \citealt{Denisenko:2012}, respectively. We also found evidence of a blended white dwarf/M-dwarf pair in the spectrum of \Gaia{} EDR3 710040763560788608, with archival sky images from SDSS confirming this scenario. With these considerations in mind, we opted for excluding these three objects from our final catalogue of polluted sources.

\subsubsection{Spectral Type Validation}  \label{sec:spectraltype_check}

Among the remaining 73 polluted white dwarf candidates, 58 were known to the MWDD and were listed in the Simbad Astronomical Database. For these known systems, we checked whether the literature spectral type classifications were consistent with our results. While the vast majority of objects were classified as white dwarfs of various types, this cross-check identified two binary systems and a subdwarf, which we also excluded from our final polluted sample: \Gaia{} EDR3 254092090595748096 ($P_{\rm{WD}}=0.90$, MWDD spectral type: DAO+BP, \citealt{Gianninas:2011}), \Gaia{} EDR3 732880265768565248 ($P_{\rm{WD}}=0.90$, MWDD spectral type: DA+M, \citealt{RebassaMansergas:2010}), and \Gaia{} EDR3 1598042809833738496 ($P_{\rm{WD}}=0.84$, MWDD spectral type: sdOB, \citealt{Eisenstein:2006}). In \autoref{fig:discard_spectype}, we present their LAMOST low-resolution spectra.

\subsubsection{Fiber Contamination}  \label{sec:contamination_check}

Next, we investigated the possibility of fiber contamination in the spectra of the remaining 70 white dwarf candidates. Similarly to the cross-match procedure described in Section \ref{sec:methodology_creation_catalog}, we employed the CDS XMatch service to identify nearby stellar objects within a radius of 3.3\arcsec{} --an angular distance comparable to the diameter of the LAMOST spectroscopic fibers. From this search, we detected three polluted white dwarf candidates with nearby stellar companions at about 1.24\arcsec{}, 2.25\arcsec{}, and 3.10\arcsec{}: \Gaia{} EDR3 203931163247581184 ($P_{\rm{WD}}=0.90$, \citealt{Greenstein:1974}), \Gaia{} EDR3 3875365174618907264 ($P_{\rm{WD}}=1.00$, \citealt{Green:1986}), and \Gaia{} EDR3 3683519503881169920 ($P_{\rm{WD}}=0.99$, \citealt{Koester:2009_2}).

As illustrated in \autoref{fig:discard_fibercontamination}, the contaminant of \Gaia{} EDR3 3875365174618907264 is of comparable brightness to the white dwarf ($G_{\rm mag}$=16.6, $G_{\rm mag, comp}$=15.6). As a consequence, we decided to exclude it from our final polluted sample. In contrast, \Gaia{} EDR3 203931163247581184 ($G_{\rm mag}$=14.8, $G_{\rm mag, comp}$=21.0) and \Gaia{} EDR3 3683519503881169920 ($G_{\rm mag}$=14.0, $G_{\rm mag, comp}$=20.1) share their LAMOST field-of-view with a significantly fainter companion, so we chose to keep them in our final catalogue of polluted white dwarfs.

\begin{figure*} 
    \begin{center}
        \includegraphics[width=0.96\linewidth]{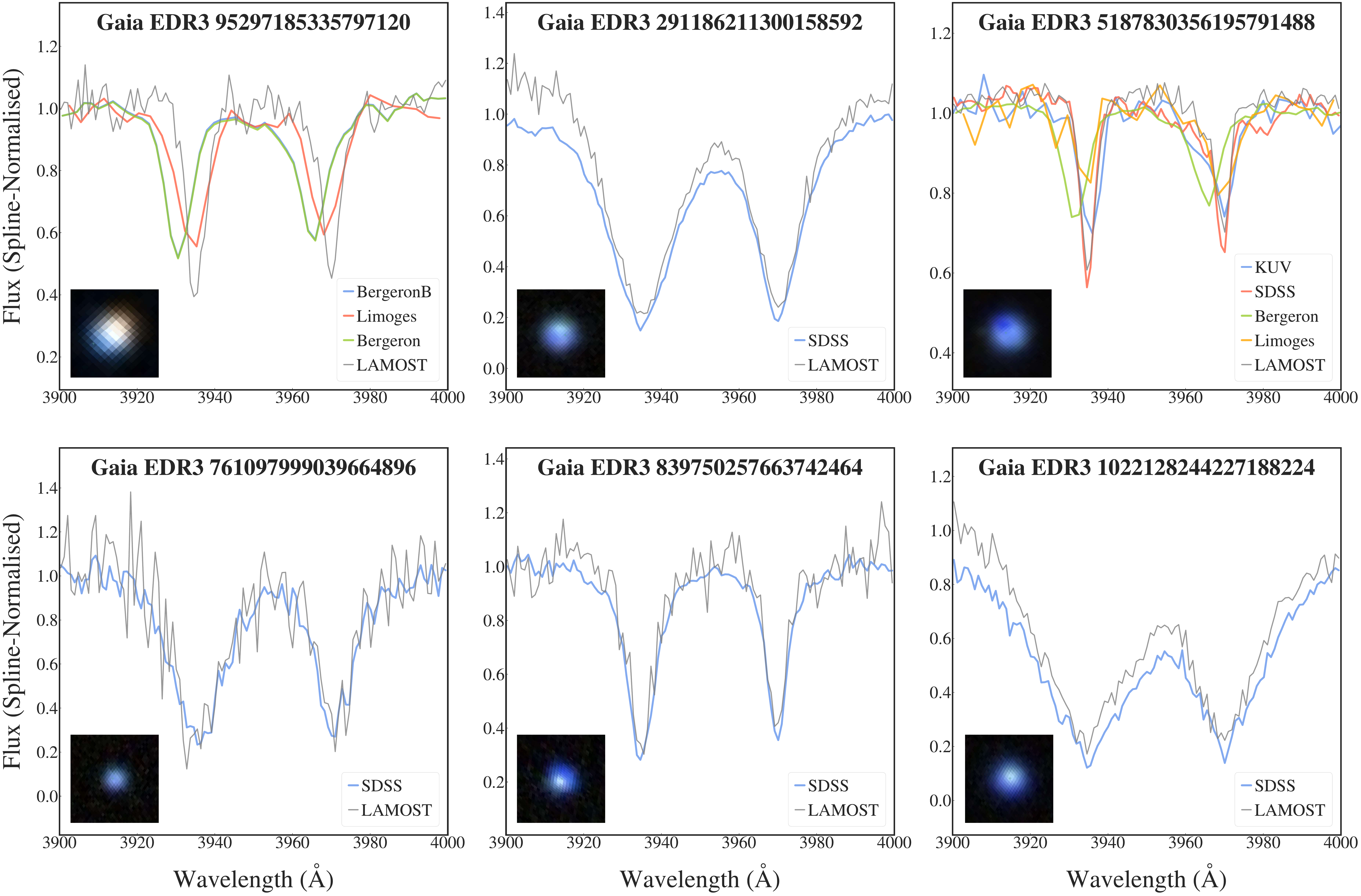}
        \caption{A selection of six ``P''/polluted LAMOST spectra (in grey) consistent with existing spectroscopic observations in the MWDD. Each panel shows the wavelength region between 3,900~\angs{} and 4,000~\angs, with the inset figures illustrating SDSS DR9 archival images of the stars' field-of-view. The legend indicates the name of the instrument or Principal Investigator to which the data belongs.} 
        \label{fig:goodmatches}
    \end{center}
\end{figure*}

\begin{figure*} 
    \begin{center}
        \includegraphics[width=1\linewidth]{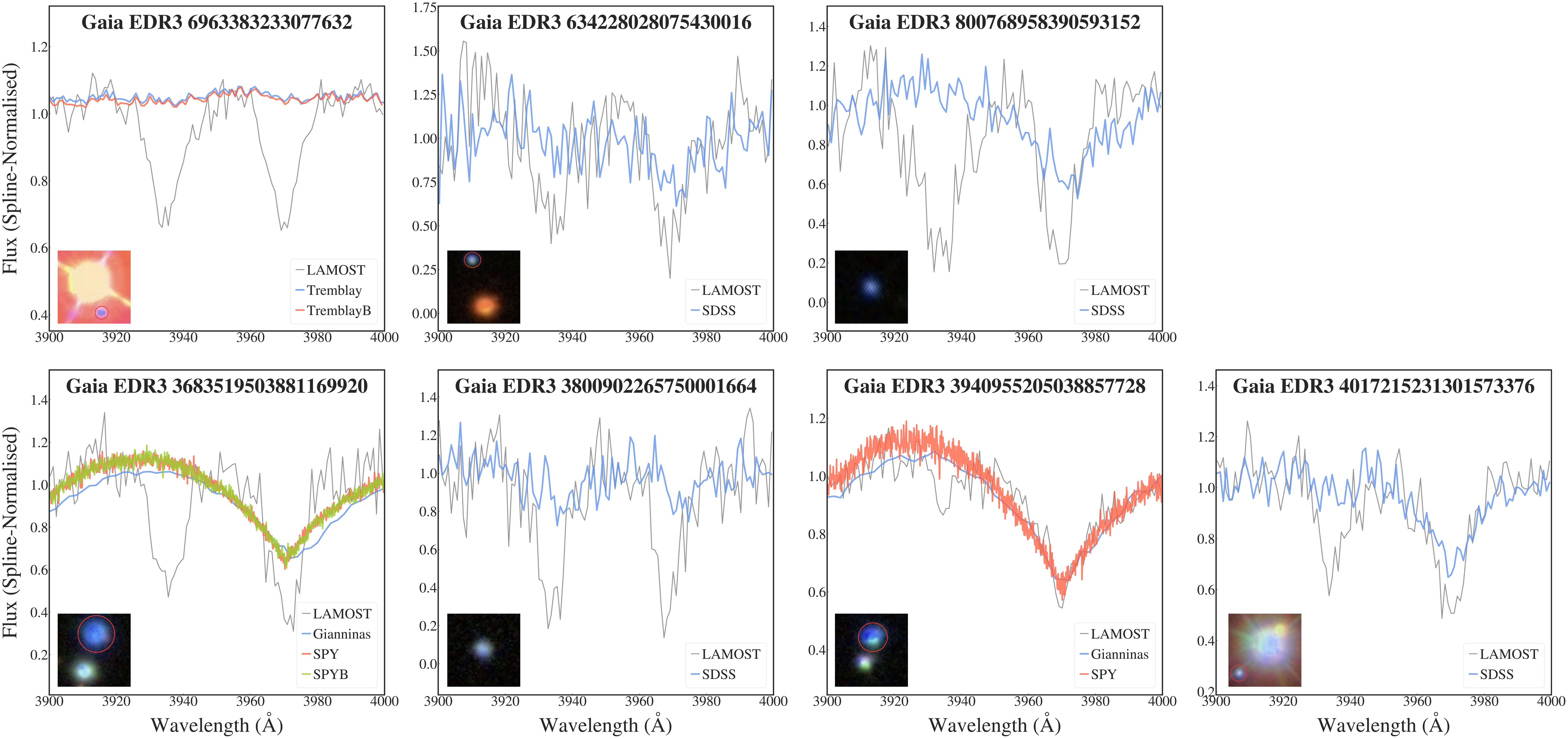}
        \caption{LAMOST spectra (in grey) of the seven``P''/polluted white dwarfs with a LAMOST spectrum inconsistent with existing MWDD spectroscopy. The inset plots show SDSS DR9 archival images of their respective field-of-view. For five stars, there is a nearby brigther companion that could be responsible for the observed discrepancies. Similarly to \autoref{fig:goodmatches}, each panel shows the wavelength region between 3,900~\angs{} and 4,000~\angs. }
        \label{fig:badmatches}
    \end{center}
\end{figure*}

\subsubsection{Comparison with Existing MWDD Spectroscopy}  \label{sec:lamost_consistency}

\begin{figure*}
    \centering
    \includegraphics[width=0.82\textwidth]{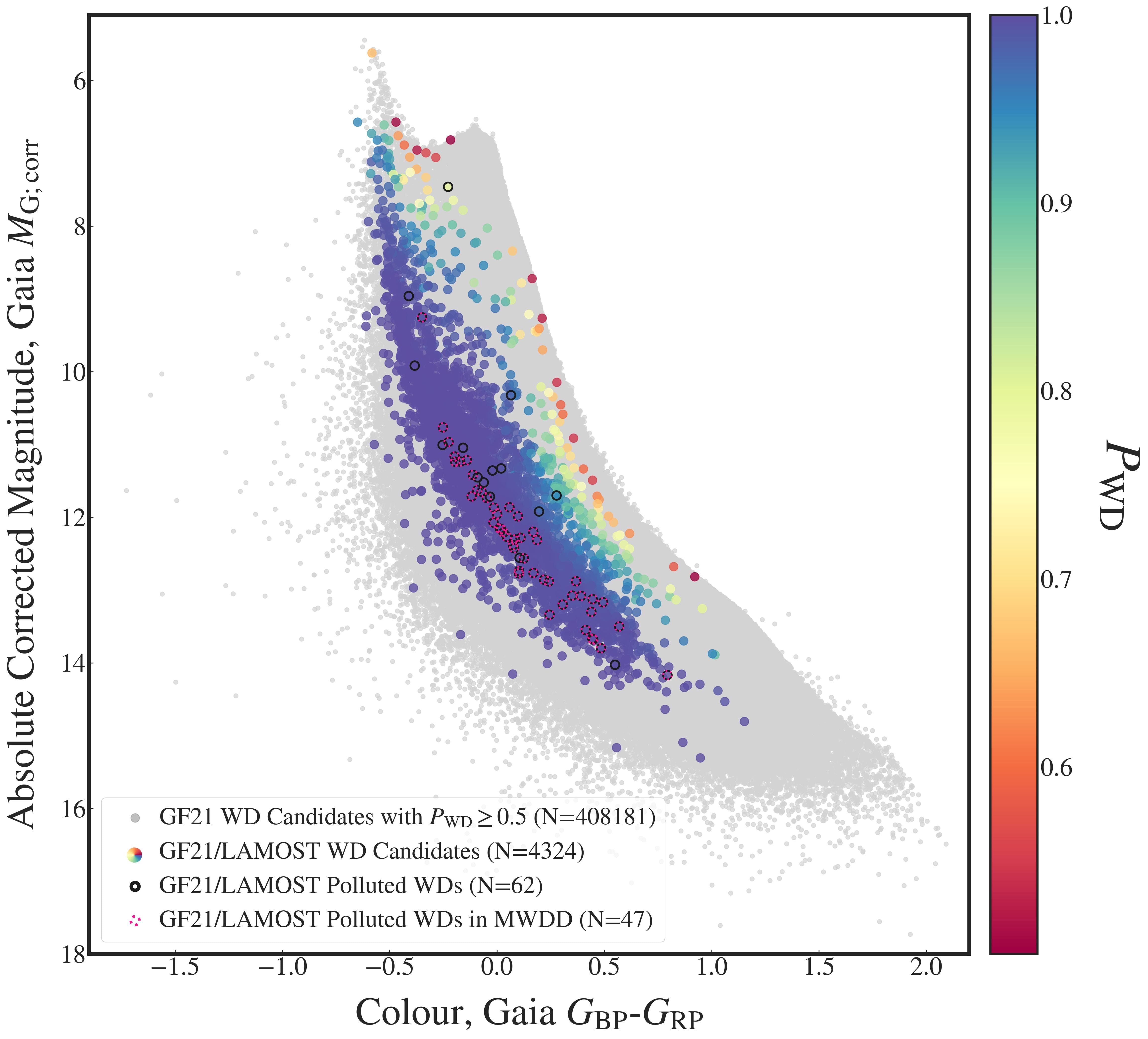}
    \caption{H-R diagram of the 4,324 GF21/LAMOST white dwarf candidates in our catalogue, with their marker colour representing their probability of being a white dwarf ($P_{\rm WD}$) as defined by GF21. We also show the 62 polluted white dwarfs identified in Section \ref{sec:final_polluted_sample} (circles with a black edge), including those already known to the MWDD (circles with both a black and red edge). The white dwarf candidates of GF21 satisfying $P_{\rm WD}\geq0.5$ (in light grey) are also included in this figure for reference purposes.}
   \label{fig:hr}
\end{figure*}

Finally, we examined the qualitative differences between the LAMOST spectra of the remaining 69 polluted white dwarf candidates and those acquired by other astronomical facilities. In particular, among the 69 stars, 54 were known to the MWDD, of which 46 had public spectroscopic observations stored in this archive --mostly from the Sloan Digital Sky Survey (SDSS; \citealt{York:2000}). For these objects, we downloaded their MWDD spectra, detrended them with basis spline using the open-source \texttt{keplersplinev2} package,\footnote{\url{https://github.com/avanderburg/keplersplinev2}.} and compared them to their LAMOST counterpart.

Our comparative study shows that the majority of polluted white dwarfs exhibit a LAMOST spectrum consistent with existing MWDD spectroscopy from other instruments --as should be expected when observing the same reference star (\autoref{fig:goodmatches}). Nonetheless, we found 7 stars with a discrepant LAMOST spectrum, as illustrated in \autoref{fig:badmatches}. For these objects, the observed inconsistencies could arise from multiple factors, such as contamination of the LAMOST spectroscopic fibers from nearby stellar companions, or cross-matching inaccuracies due to small coordinate differences in the LAMOST or other databases. Based on archival images from SDSS DR9, these hypotheses could be plausible for 5 of the 7 white dwarf candidates with discrepant observations (\Gaia{} EDR3 6963383233077632, 634228028075430016, 3683519503881169920, 3940955205038857728, and 4017215231301573376), all of which have a brighter companion within 13\arcsec{} (see inset plots in \autoref{fig:badmatches}). For the remaining 2 stars (\Gaia{} EDR3 800768958390593152 and 3800902265750001664), MAST visual images of their field-of-view show no evidence of nearby  objects that may have been ommitted in the \Gaia{} database. As a result, the cause of the observed discrepancies is more unclear, and new data will be needed to validate their LAMOST spectra. In the remaining sections of the paper, we adopted a conservative approach and decided to exclude the 7 discrepant stars from our final polluted sample.

\begin{figure}
    \centering
    \includegraphics[width=0.96\columnwidth]{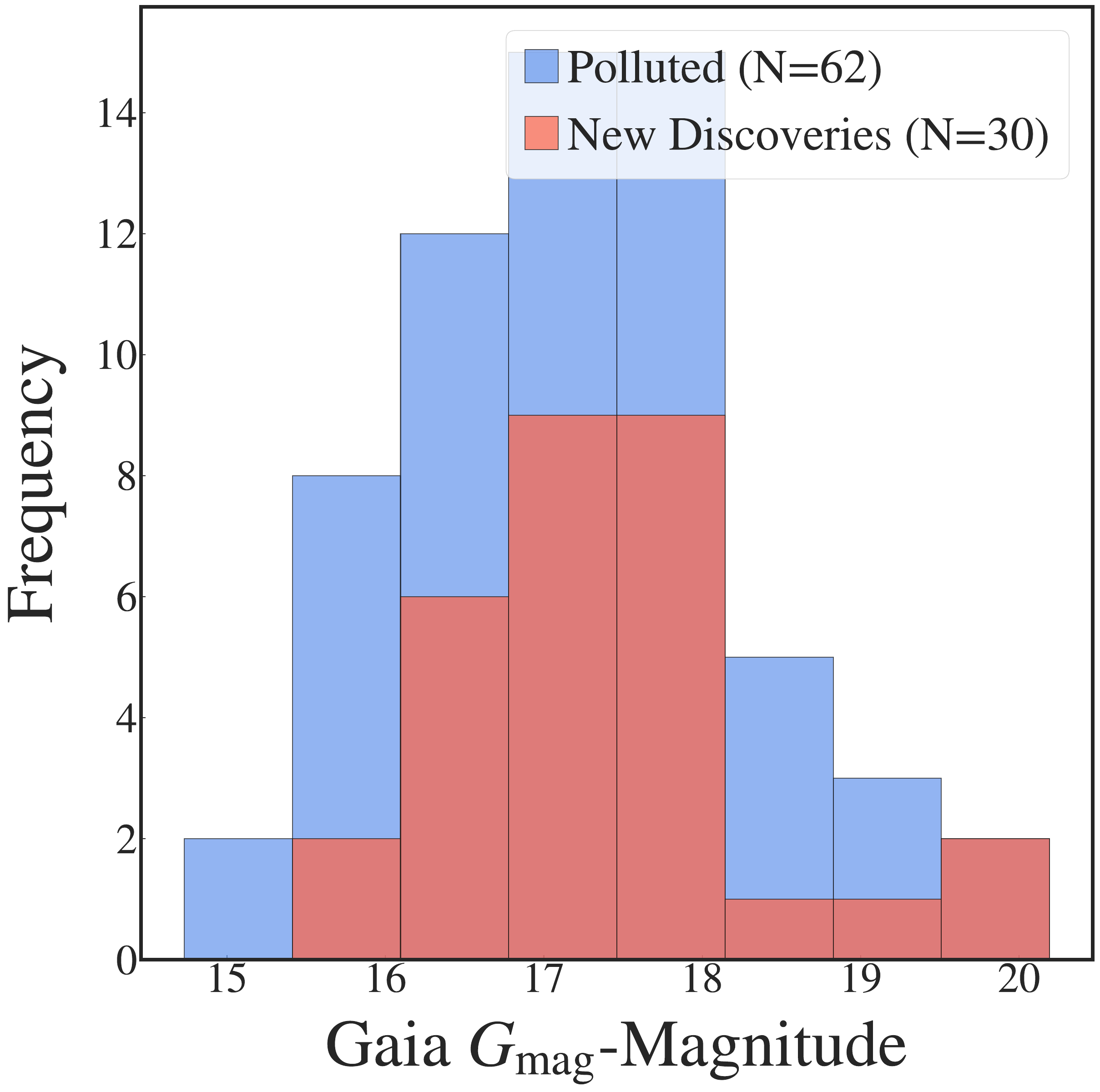}
    \caption{Distribution of \Gaia{} $G_{\rm mag}$-magnitudes for our sample of 62 polluted white dwarfs (in light blue), including 30 systems with previously unknown calcium signatures (in red). Magnitude values are taken from the GF21 database and can be found in the ``\textsc{G\_mag}'' column of our catalogue.}
   \label{fig:gmag_histogram}
\end{figure}

\subsection{Final Catalogue of Polluted White Dwarfs}   \label{sec:final_polluted_sample}

Our quantitative study of calcium pollution yielded a preliminary list of 76 polluted white dwarf candidates with calcium significances greater than $\alpha\geq1.90$ at about 3,934~\angs{} and 3,969.5~\angs. We then refined and validated this list by visually inspecting their LAMOST spectra and performing multiple literature cross-checks. Through this vetting process, we discarded 14 objects (see Table \ref{tab:discarded_from_catalog}) and obtained a final sample of 62 polluted white dwarfs --or about $\sim$1.43$\%$ relative to the entire GF21/LAMOST catalogue.\footnote{These polluted white dwarfs are flagged with the letter ``Y''  in the \textsc{CaPollution} column of our catalogue (see Table \ref{tab:catalog_format}). The remaining non-polluted stars are classified with an ``N.''} In Table \ref{tab:props_polluted}, we present their main astrophysical properties, including the statistical significances of their H8, Ca II, and combined Ca II/H$\epsilon$ lines, and our own spectral classification, with the order of the letters reflecting the strength of their dominant absorption lines.

\begin{table*}
    \centering
    \caption{Final GF21/LAMOST catalogue of 62 polluted white dwarfs, including 30 of them with previously unknown calcium pollution (in blue, see Section \ref{sec:new_discoveries}). The \textsc{obsid} keyword is the LAMOST identifier of the star. The \textsc{R.A.} and \textsc{Dec.} coordinates are taken from GF21 and are given relative to ICRS J2016. The \textsc{$\alpha_{\mathrm{H8}}$},  $\alpha_{\mathrm{CaII}}$, and $\alpha_{\mathrm{CaII,H\epsilon}}$ columns provide, respectively, the statistical significances of the H8, Ca II, and blended Ca II/H$\epsilon$ lines (see Section \ref{sec:lorentz_classification}). The first and second \textsc{SpC} columns show, respectively, our spectral classification based on the strength of the observed absorption lines, and the default spectral type in the MWDD as of September 2023; if the two are different, we indicate the preferred classification with a dagger symbol ($\dagger$) based on the quality of the competing spectra. The logarithmic number abundance ratios for hydrogen and calcium are reported in base 10.} 
    \label{tab:props_polluted}
    \addtolength{\tabcolsep}{-5pt}   
    \renewcommand{\arraystretch}{1.2}
    \begin{tabular}{|c|ccccc|cccc|cccccc|}
    \hline
    & \multicolumn{5}{c|}{\it Main Properties from LAMOST and GF21} & \multicolumn{4}{c|}{\it This Work}  &  \multicolumn{6}{c|}{\it MWDD}   \\  
    \hline
    &  \Gaia{} EDR3          & obsID     &  R.A.    & Dec.       & $G_{\rm mag}$ & $\alpha_{\mathrm{H8}}$ & $\alpha_{\mathrm{CaII}}$ & $\alpha_{\mathrm{CaII,H\epsilon}}$ &  SpC & SpC & \Teff & \logg  & $\log(\rm H/He)$ & $\log(\rm Ca/H)$& $\log(\rm Ca/He)$  \\
    \hdashline[0.2pt/1.5pt]
    \small{No.}  &           &      &  \small{[deg]}    & \small{[deg]}      & &  &  &  &  &   & \small{[K]} & \small{[cgs]} &  &  &  \\
    \hline
    \hline
    1  & 1022128244227188224           &  421304157 &  144.93 &  55.85 & 16.74 &   0.11  &  11.02 & 15.88  &  DZA$^{\dagger}$  &  DZ                  & 8888$\pm$40   & 8.01$\pm$0.01 &  -4.00  &  -     & -8.48  \\       
    2  & 1032748903780398592           &  604516136 &  121.30 &  54.10 & 17.70 &   3.33  &  2.71  & 2.10   &  DBZA             &  DBZA                & 15250         & 8.78          &  -5.70  &  -     & -9.27  \\       
    3  & 1193520666521113344           &  565711079 &  233.77 &  12.80 & 15.59 &   1.21  &  3.38  & 5.04   &  DZ               &  DZ                  & 5969$\pm$10   & 8.17$\pm$0.00 &  -3.00  &  -     & -8.48  \\       
    4  & 1196531988354226560           &  558410149 &  238.54 &  17.36 & 17.26 &   7.11  &  7.98  & 5.53   &  DBAZ             &  DBAZ                & 16252         & 8.08          &  -4.20  &  -     & -6.47  \\       
    5  & 1202552914026910976           &  558410130 &  238.62 &  17.60 & 17.45 &   0.06  &  1.97  & 2.21   &  DZ               &  DZ                  & 6794$\pm$25   & 8.21$\pm$0.01 &  -4.00  &  -     & -8.59  \\       
    6  & 1336442472164656000           &  914109128 &  262.89 &  37.09 & 16.03 &   1.75  &  5.09  & 5.03   &  DAZ              &  DAZB$^{\dagger}$    & 9825$\pm$59   & 7.99$\pm$0.01 &  -1.5   &  -5.1  & -7.00  \\       
    7  & \blue{1411668854417599104}    &  743009135 &  246.87 &  49.42 & 18.89 &   0.06  &  1.93  & 2.34   &  DAZ$^{\dagger}$  &  DA                  & 7930          & 8.05          &  -      &  -     & -      \\       
    8  & \blue{1480890365586927744}    &  901610055 &  214.88 &  35.76 & 17.76 &   0.24  &  2.56  & 3.15   &  DAZ$^{\dagger}$  &  -                   & -             & -             &  -      &  -     & -      \\       
    9  & 1527809271227078272           &  342506137 &  194.47 &  42.88 & 17.23 &   5.26  &  2.15  & 2.06   &  DBAZ             &  DBAZ                & 16015         & 8.07          &  -5.58  &  -     & <-7.03 \\       
    10 & \blue{1565012935076040832}    &  564901178 &  202.66 &  54.76 & 16.72 &   1.67  &  2.26  & 3.77   &  DAZ$^{\dagger}$  &  DA                  & 7583$\pm$20   & 8.12$\pm$0.01 &  -      &  -     & -      \\       
    11 & \blue{1982029516212384384}    &  843010181 &  338.09 &  43.82 & 17.18 &   0.78  &  5.55  & 6.21   &  DAZ$^{\dagger}$  &  -                   & -             & -             &  -      &  -     & -      \\       
    12 & \blue{200924312484080256}     &  302715031 &  75.50  &  41.05 & 17.52 &   4.19  &  2.10  & 5.41   &  DAZ$^{\dagger}$  &  -                   & -             & -             &  -      &  -     & -      \\       
    13 & 203931163247581184$^{\rm co}$ &  295304028 &  69.66  &  41.16 & 14.82 &   29.61 &  5.56  & 8.38   &  DBZA             &  DBZA                & 5680$\pm$1633 & 8.04          &  -4.34  &  -     & -8.51  \\  
    14 & 234469931207274112            &  284114213 &  49.91  &  36.51 & 16.33 &   0.12  &  2.67  & 2.74   &  DZ               &  DZ                  & 6467$\pm$29   & 8.15$\pm$0.01 &  -4.00  &  -     & -9.26  \\       
    15 & \blue{2503322616188632320}    &  77505159  &  37.86  &  2.76  & 20.19 &   1.03  &  5.50  & 7.51   &  DAZ$^{\dagger}$  &  -                   & -             & -             &  -      &  -     & -      \\       
    16 & \blue{2564020159166337024}    &  472002174 &  18.04  &  4.92  & 16.82 &   4.05  &  8.44  & 7.75   &  DAZ$^{\dagger}$  &  -                   & -             & -             &  -      &  -     & -      \\       
    17 & 2564945432560219008           &  354812027 &  23.10  &  5.49  & 18.21 &   1.20  &  5.36  & 4.57   &  DAZ              &  DAZ                 & 7355$\pm$41   & 7.98$\pm$0.02 &  -      &  -7.44 & -      \\       
    18 & 2730508416002618752           &  157014186 &  337.01 &  12.13 & 16.20 &   0.03  &  2.03  & 2.10   &  DZ               &  DZ                  & 6885$\pm$18   & 8.1$\pm$0.0   &  -4.00  &  -     & -9.73  \\       
    19 & \blue{27578539058961280}      &  368602177 &  45.14  &  11.85 & 17.71 &   0.87  &  4.20  & 4.38   &  DAZ$^{\dagger}$  &  -                   & -             & -             &  -      &  -     & -      \\       
    20 & \blue{2759588063311504768}    &  355001148 &  355.86 &  8.62  & 17.27 &   2.31  &  2.69  & 4.20   &  DAZ$^{\dagger}$  &  DA                  & 6836$\pm$39   & 7.98$\pm$0.01 &  -      &  -     & -      \\       
    21 & 291186211300158592            &  277801215 &  27.39  &  24.01 & 15.97 &   0.10  &  18.89 & 19.85  &  DAZ$^{\dagger}$  &  DZ                  & 8762$\pm$28   & 8.09$\pm$0.00 &  -3.00  &  -     & -8.45  \\       
    22 & \blue{295237976073772672}     &  202601236 &  23.07  &  26.12 & 19.84 &   0.63  &  12.78 & 18.83  &  DAZ$^{\dagger}$  &  -                   & -             & -             &  -      &  -     & -      \\       
    23 & 320029150076023808            &  201401159 &  20.41  &  34.68 & 16.61 &   0.23  &  7.56  & 5.28   &  DZ               &  DZ                  & 6798$\pm$28   & 8.09$\pm$0.01 &  -4.00  &  -     & -10.85 \\       
    24 & \blue{321093335597030400}     &  201402181 &  18.68  &  35.41 & 18.70 &   1.81  &  4.18  & 2.49   &  DAZ$^{\dagger}$  &  DA                  & -             & -             &  -      &  -     & -      \\       
    25 & 322324719900568448            &  284009139 &  22.22  &  36.88 & 17.32 &   4.79  &  8.64  & 5.13   &  DBZ              &  DBZ                 & -             & -             &  -      &  -     & -      \\       
    26 & 3253852143823041792           &  372704024 &  64.63  &  -2.32 & 17.17 &   1.52  &  2.46  & 2.61   &  DAZ$^{\dagger}$  &  -                   & -             & -             &  -      &  -     & -      \\       
    27 & \blue{333392060349558784}     &  778504097 &  39.25  &  36.75 & 17.68 &   1.84  &  2.50  & 3.37   &  DAZ$^{\dagger}$  &  -                   & -             & -             &  -      &  -     & -      \\       
    28 & \blue{3351139990665573120}    &  387301118 &  101.49 &  11.21 & 17.65 &   5.09  &  3.98  & 2.50   &  DBAZ$^{\dagger}$ &  DBA                 & -             & -             &  -      &  -     & -      \\       
    29 & \blue{3360183783038606336}    &  422409198 &  106.37 &  16.2  & 16.60 &   5.39  &  7.20  & 8.57   &  DAZ$^{\dagger}$  &  CND                 & 9516          & 7.98          &  -      &  -     & -      \\       
    30 & \blue{340886198462842112}     &  854612056 &  39.72  &  44.55 & 16.40 &   2.28  &  5.22  & 7.13   &  DAZ$^{\dagger}$  &  -                   & -             & -             &  -      &  -     & -      \\       
    31 & \blue{3433524747667276544}    &  105809172 &  95.05  &  27.93 & 17.38 &   0.17  &  1.96  & 2.57   &  DZ$^{\dagger}$   &  CND                 & 8672          & 8.35          &  -      &  -     & -      \\       
    32 & \blue{3440707857129932672}    &  185908034 &  95.26  &  35.07 & 17.19 &   2.79  &  2.22  & 4.06   &  DAZ$^{\dagger}$  &  CND                 & 7982          & 7.96          &  -      &  -     & -      \\       
    33 & 3583181371265430656           &  213306229 &  186.95 &  -8.24 & 14.73 &   0.92  &  12.35 & 14.93  &  DZA              &  DBAZ$^{\dagger}$                & 10800         & 8.00          &  -4.05  &  -     & -8.06  \\       
    34 & \blue{3669065354086975872}    &  812815099 &  217.70 &  4.64  & 17.01 &   4.15  &  3.34  & 7.94   &  DAZ$^{\dagger}$  &  DA                  & 7980$\pm$37   & 8.08$\pm$0.01 &  -      &  -     & -      \\       
    35 & \blue{367949367212923392}     &  353915195 &  12.30  &  38.69 & 17.13 &   8.38  &  2.82  & 12.81  &  DAZ$^{\dagger}$  &  DA                  & 9830$\pm$53   & 7.98$\pm$0.02 &  -      &  -     & -      \\       
    36 & 3808050667584060672           &  438216197 &  164.01 &  1.47  & 17.66 &   0.01  &  2.70  & 3.06   &  DZ               &  DZ                  & 10278$\pm$378 & 8.07          &  -4.92  &  -     & -8.99  \\       
    37 & 3818299730867320704           &  631804173 &  166.16 &  7.19  & 18.15 &   2.31  &  6.30  & 4.35   &  DZ               &  DZ                  & 16488         & 9.34          &  -5.08  &  -     & -9.58  \\       
    38 & 3836479227914477184           &  316604156 &  150.66 &  3.22  & 18.96 &   0.01  &  3.12  & 2.58   &  DZ               &  DZ                  & 9997$\pm$333  & 8.06          &  <-5.87 &  -     & -8.89  \\       
    39 & \blue{3878793490528218752}    &  868302211 &  148.03 &  9.16  & 17.12 &   6.59  &  2.49  & 2.36   &  DBAZ$^{\dagger}$ &  DB                  & 18762         & 8.21          &  -4.58  &  -     & <-5.09 \\       
    40 & 4002914643768684288           &  629804063 &  180.83 &  24.67 & 16.70 &   8.35  &  2.51  & 2.13   &  DBAZ             &  DBZA                & 15300         & 8.60          &  -4.91  &  -     & -9.40  \\       
    41 & 4014824794238339840           &  231104199 &  182.72 &  31.62 & 18.89 &   0.02  &  3.86  & 3.65   &  DZ               &  DZ                  & 9121$\pm$275  & 8.12          &  <-5.57 &  -     & -9.18  \\       
    42 & \blue{4502205519647039616}    &  746414228 &  270.78 &  16.58 & 17.75 &   0.03  &  4.68  & 5.44   &  DZ$^{\dagger}$   &  CND                 & 9173          & 8.27          &  -      &  -     & -      \\       
    43 & \blue{480570075502703488}     &  613507047 &  85.00  &  66.51 & 16.43 &   1.04  &  2.92  & 2.83   &  DAZ$^{\dagger}$  &  -                   & -             & -             &  -      &  -     & -      \\       
    44 & 5187830356195791488           &  837313013 &  45.72  &  -1.14 & 15.49 &   18.27 &  24.84 & 15.83  &  DBZA             &  DBZA                & 14430         & 7.70          &  -6.25  &  -     & -7.58  \\       
    45 & \blue{642549544391197440}     &  713109166 &  151.05 &  24.65 & 16.47 &   3.69  &  2.03  & 7.26   &  DAZ$^{\dagger}$  &  CND                 & 10919$\pm$66  & 7.97          &  -      &  -     & -      \\       
    \vdots &                           &  \vdots    &         & \vdots &       & \vdots  &        & \vdots &                   &   \vdots &               & \vdots        &         & \vdots &        \\ 
    \end{tabular}    
\end{table*}
\addtolength{\tabcolsep}{1pt}

\begin{table*}
    \centering
    \contcaption{Table \ref{tab:props_polluted}. }
    \addtolength{\tabcolsep}{-5.80pt}   
    \renewcommand{\arraystretch}{1.2}
    \begin{tabular}{|c|ccccc|cccc|cccccc|}
    \hline
    & \multicolumn{5}{c|}{\it Main Properties from LAMOST and GF21} & \multicolumn{4}{c|}{\it This Work}  &  \multicolumn{6}{c|}{\it MWDD}   \\  
    \hline
    &  \Gaia{} EDR3          & obsID     &  R.A.    & Dec.       & $G_{\rm mag}$ & $\alpha_{\mathrm{H8}}$ & $\alpha_{\mathrm{CaII}}$ & $\alpha_{\mathrm{CaII,H\epsilon}}$ &  SpC & SpC & \Teff & \logg  & $\log(\rm H/He)$ & $\log(\rm Ca/H)$& $\log(\rm Ca/He)$  \\
    \hdashline[0.2pt/1.5pt]
    \small{No.}  &           &      &  \small{[deg]}    & \small{[deg]}      & &  &  &  &  &   & \small{[K]} & \small{[cgs]} &  &  &  \\
    \hline
    \hline
    \vdots &                          &  \vdots     &         & \vdots &       & \vdots  &        & \vdots &                   & \vdots   &               & \vdots        &         & \vdots &       \\ 
    46 & 678517387234326528            &  321904024 &  128.59 &  24.37 & 18.55 &   1.12  &  4.32  & 3.99   &  DZ               &  DZ                  & 11590$\pm$456 & 8.04          &  -5.24  &  -     & -8.18  \\       
    47 & \blue{689352219629097856}     &  601709080 &  131.41 &  22.96 & 15.88 &   15.87 &  6.16  & 5.16   &  DBZ$^{\dagger}$  &  DB                  & 19432         & 8.15          &  -5.16  &  -     & <-5.48 \\       
    48 & \blue{716504796716020352}     &  483112174 &  132.29 &  34.50 & 15.57 &   4.69  &  2.68  & 9.73   &  DAZ$^{\dagger}$  &  DA                  & 7548$\pm$22   & 7.98$\pm$0.01 &  -      &  -9.66 & -      \\ 
    49 & 717762977319125632            &  331801189 &  132.21 &  35.82 & 18.06 &   0.53  &  4.20  & 2.69   &  DZ               &  DZ                  & 7860$\pm$63   & 8.18$\pm$0.02 &  -4.00  &  -     & -10.45 \\       
    50 & 754220107131130624            &  803413167 &  154.66 &  37.45 & 17.76 &   0.02  &  4.24  & 3.63   &  DZA$^{\dagger}$  &  DZ                  & 10543$\pm$361 & 7.98          &  -3.62  &  -     & -7.93  \\       
    51 & 761097999039664896            &  321407168 &  171.44 &  38.39 & 17.91 &   0.21  &  5.35  & 5.03   &  DZ               &  DZ                  & 10437$\pm$330 & 8.17          &  <-5.88 &  -     & -8.17  \\       
    52 & \blue{762684864901596928}     &  620008163 &  164.51 &  36.20 & 17.84 &   2.95  &  2.21  & 3.55   &  DAZ$^{\dagger}$  &  DA                  & 22839         & 7.26          &  -      &  -     & -      \\       
    53 & \blue{829084990579067264}     &  34613120  &  158.28 &  43.48 & 16.60 &   0.24  &  13.09 & 19.11  &  DAZ$^{\dagger}$  &  -                   & -             & -             &  -      &  -     & -      \\       
    54 & 839750257663742464            &  42502046  &  170.15 &  52.96 & 16.75 &   0.18  &  8.20  & 7.04   &  DZA$^{\dagger}$  &  DZ                  & 10754         & 8.20          &  -      &  -     & -      \\       
    55 & 866574493091351936            &  174504020 &  113.44 &  23.26 & 17.45 &   0.04  &  5.28  & 2.99   &  DZ               &  DZ                  & 8787$\pm$64   & 8.06$\pm$0.02 &  -4.20  &  -     & -      \\       
    56 & 909933248799545344            &  722703026 &  130.48 &  37.39 & 17.78 &   2.85  &  6.92  & 2.17   &  DBZ$^{\dagger}$  &  DZ                  & 13806         & 8.19          &  <-6.52 &  -     & -7.95  \\       
    57 & 912415155781038464            &  421907104 &  132.30 &  40.61 & 18.18 &   0.38  &  4.00  & 2.23   &  DZ               &  DZ                  & 9388$\pm$295  & 8.20          &  <-5.69 &  -     & -10.12 \\       
    58 & \blue{917141857485288704}     &  793608149 &  130.04 &  44.36 & 17.81 &   8.31  &  16.42 & 21.47  &  DAZ$^{\dagger}$  &  -                   & -             & -             &  -      &  -     & -      \\       
    59 & 950361883331847424            &  296115106 &  103.58 &  39.64 & 15.92 &   0.23  &  2.60  & 3.06   &  DZ               &  DZ                  & 9673$\pm$46   & 8.02$\pm$0.01 &  -3.00  &  -     & -8.88  \\       
    60 & 95297185335797120             &  378505204 &  27.24  &  19.04 & 15.54 &   2.11  &  17.16 & 13.9   &  DAZB             &  DAZB                & 11500         & 8.00          &  -2.89  &  -     & -8.70  \\       
    61 & \blue{972551088836290816}     &  632707021 &  91.30  &  52.74 & 17.31 &   5.21  &  6.83  & 4.03   &  DABZ$^{\dagger}$ &  -                   & -             & -             &  -      &  -     & -      \\       
    62 & \blue{994846611963848704}     &  853011165 &  100.24 &  55.45 & 17.71 &   2.21  &  2.94  & 2.67   &  DAZ$^{\dagger}$  &  -                   & -             & -             &  -      &  -     & -      \\    
    \hline
    \end{tabular} 
    \vspace{2pt}
    \begin{quote}
    \hspace{0.4cm}  [$\rm\textit{co}$]: Contaminant within 1.5\arcsec{} of the white dwarf. \\
    \hspace{0.52cm} [$\dagger$]: Preferred spectral classification. Disagreements with the MWDD can be attributed to two potential causes: LAMOST allowing us to detect more elements than before, or conversely, the higher S/N of existing MWDD spectra enabling the detction of more elements than in the LAMOST data. \\
     \end{quote}
\end{table*}
\addtolength{\tabcolsep}{1pt}

\subsubsection{Identification of New Polluted White Dwarfs}\label{sec:new_discoveries}

Next, we set out to identify those stars with previously unknown signs of calcium pollution. For this task, we defined two types of objects as potential ``new discoveries,'' namely: those not listed in the MWDD as January 2023, and those included in the MWDD but not classified as polluted. In both cases, our definition relied on the MWDD as the largest and most comprehensive public database of white dwarfs known to date.

To begin our search, we cross-matched the \Gaia{} EDR3 names of the 62 polluted sources with those in the MWDD, obtaining 47 matches and 15 unknown objects (see \autoref{fig:flowchart}). Among the successful matches, 31 were listed as polluted in the MWDD (i.e. ``DZ-'' spectral types), and 16 had been classified as non-polluted --primarily falling under the ``DA'' and ``DB'' spectral types. Taking these numbers into account --in particular, the sample of 16 stars with no apparent pollution, as well as the 15 objects unknown to the MWDD--, our cross-match routine suggested the discovery of 31 potentially new polluted white dwarfs. To validate these findings, we performed a literature search with Simbad, identifying one object with known calcium pollution (\Gaia{} EDR3 3253852143823041792, spectral type: DBAZ, \citealt{Kong:2019}). The remaining 30 white dwarfs represent new discoveries, with LAMOST providing their first-ever public spectroscopic observations to date.

In \autoref{fig:hr}, we present an H-R diagram of the 4,324 white dwarf candidates in our catalogue. As expected due to the $P_{\rm WD}\geq0.50$ criterion imposed in Section \ref{sec:methodology_creation_catalog}, the majority of GF21/LAMOST systems are high-probability white dwarf candidates (i.e. $P_{\rm WD}\geq0.75$), with only 1 star satisfying $0.50\leq P_{\rm WD}<0.75$  (\Gaia{} EDR3 480570075502703488). In \autoref{fig:gmag_histogram}, we also illustrate the distribution of \Gaia{} $G_{\rm mag}$-magnitudes for the 62 polluted white dwarfs, including the 30 new discoveries. As this histogram shows, the mean brightness of the full polluted catalogue is about $\bar{G}_{\rm mag}$=17.2, which is about 20$\%$ brighter than the typical magnitude of the 30 previously unknown polluted white dwarfs ($\bar{G}_{\rm mag}$=17.4). In both cases, however, there are a handful of brighter objects ($G_{\rm mag}\leq16.5$) that could be particularly suitable for follow-up characterisation work at higher spectral resolution; we elaborate more on this subject in Section \ref{sec:observational_priorities}.

\subsubsection{Search for Other Heavy Elements} \label{sec:other_elements} 

While the number of polluted white dwarfs  identified in this work appears to be relatively low ($\sim1.43\%$), it is likely that this result represents a conservative estimate. Indeed, given the low S/N and resolving power of LAMOST, it is possible that our Lorentzian fitting algorithm was not sensitive enough to the weakest signs of photospheric pollution. Nonetheless, the presence of calcium in 62 systems is promising, particularly for those sources with deep absorption lines (e.g. \Gaia{} EDR3 5187830356195791488 in \autoref{fig:lorentzian_fit}). In the future, high-resolution spectroscopy of some of these white dwarfs may enable the detection of more heavy elements, especially for He-atmosphere white dwarfs, which tend to exhibit richer spectral features due to their increased transparency \citep{Dufour:2012, Klein:2021, Saumon:2022}. In Section \ref{sec:observational_priorities}, we offer recommendations for target selection based on multiple physical and observational considerations, including the white dwarf's effective temperature, brightness, and amount of calcium pollution.

To help prioritise targets for future spectroscopic campaigns, we complemented our analysis of calcium pollution with a visual assessment of the presence of magnesium and iron in the 62 polluted white dwarf spectra. More specifically, we searched for traces of Mg I, Mg II, Fe I, and Fe II across the full wavelength range of LAMOST,\footnote{To perform our search, we examined the following air wavelengths \citep{NIST_ASD}: (i) \textit{for Mg I}: 3,829.36~\angs{}, 3,832.30~\angs{}, 3,838.29~\angs{}, 4,702.99~\angs{}, 5,167.32~\angs{}, 5,172.68~\angs{}, 5,183.60~\angs{}, 5,528.41~\angs{}, 5,711.09~\angs{}, 8,806.76~\angs{}; (ii) \textit{for Mg II}: 4,390.56~\angs{}, 4,481.13~\angs{}, 6,346.74~\angs{}, 7,877.051~\angs{}, 7,896.37~\angs{}, 8,213.99~\angs{}, 8,234.64~\angs{}, 8,734.99~\angs{}, 8,745.66~\angs{}, 8,824.32~\angs{}, 8,835.08~\angs{}; (iii) \textit{for Fe I}:  3,440.61~\angs{}, 3,475.45~\angs{}, 3,490.57~\angs{}, 3,705.57~\angs{}, 3,722.56~\angs{}, 3,745.56~\angs{}, 3,748.26~\angs{}, 3,878.57~\angs{}; and (iv) \textit{for Fe II}: 5,001.95~\angs{}, 5,227.49~\angs{}, 5,260.254~\angs{}.} identifying 11 observations with both Ca and Mg pollution, and none with convincing Fe lines. Although a more robust and comprehensive analysis is needed to validate our possible magnesium detections, we include the results of our visual search in the ``\textsc{mg\_detection}'' column of our catalogue.

\begin{figure*} 
    \begin{center}
        \includegraphics[width=0.9\linewidth]{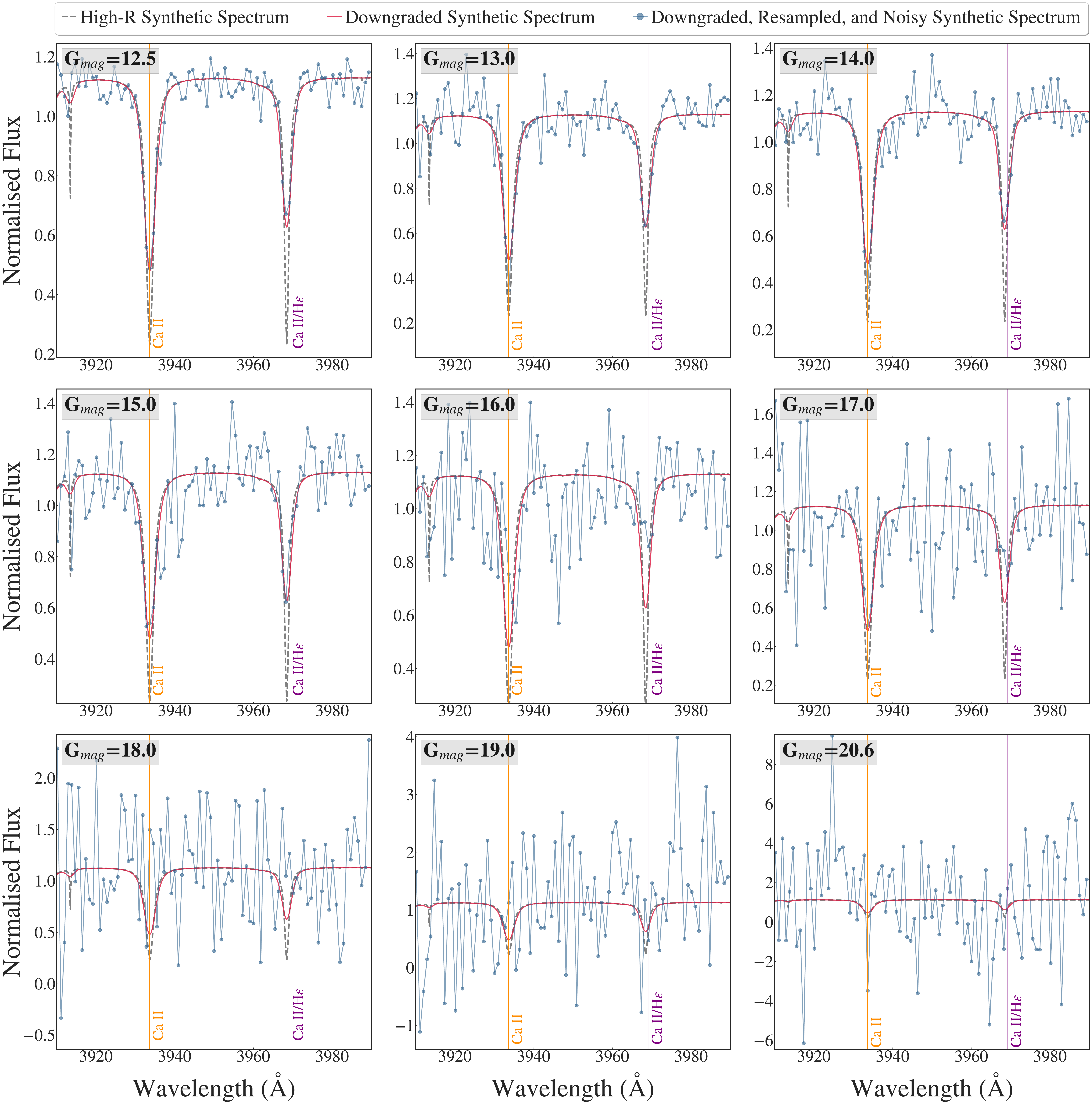}
        \caption{Effects of downgrading, resampling, and injecting a representative amount of LAMOST Gaussian noise into the high-resolution synthetic spectrum of a He-rich white dwarf with \Teff=11,066~K, \logg=7.30 cgs, and \logCaHe=-9.50. The orange and purple lines show the Ca II and blended Ca II/H$\epsilon$ absorption features at about 3,934~\angs{} and 3,969.5~\angs{}, respectively. Each panel shows a reference \Gaia{} ${G}_{\rm mag}$-magnitude between 12.5 and 20.6.}
        \label{fig:montreal}
    \end{center}
\end{figure*}

\begin{figure*} 
    \begin{center}
        \includegraphics[width=0.92\linewidth]{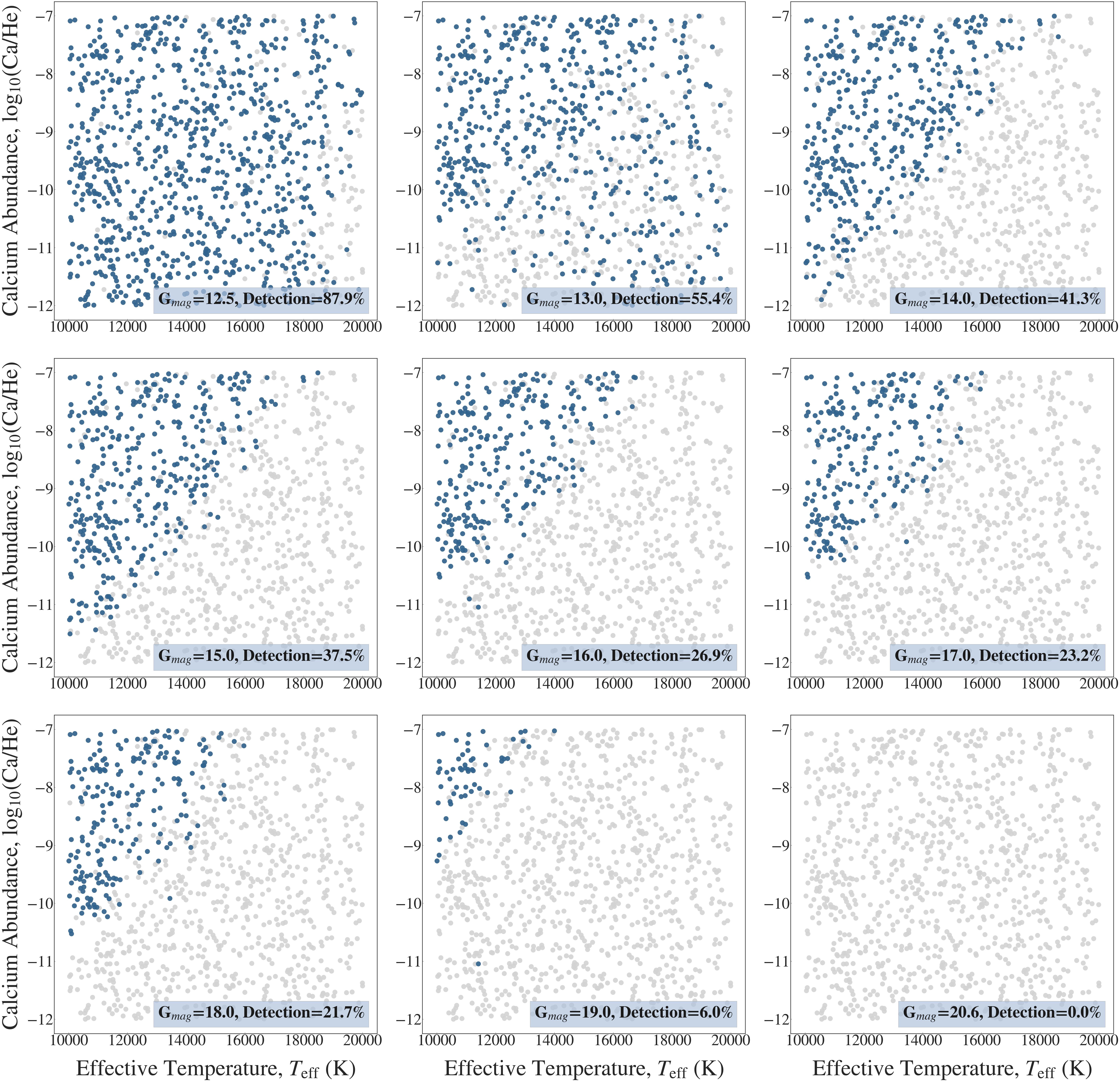}
        \caption{Detection limits of the LAMOST spectrograph for He-rich white dwarfs as a function of synthetic logarithmic calcium abundance and effective temperature. }
        \label{fig:ca_threshold}
    \end{center}
\end{figure*}

\begin{table}
    \centering
    \caption{White dwarf candidates from the full GF21/LAMOST catalogue with a flux error representative of multiple \Gaia{} ${G}_{\rm mag}$ reference magnitudes. The third and fourth columns provide, respectively, their true ${G}_{\rm mag}$ and the median of their flux error between 3,800~\angs{} and 4,000~\angs. \label{tab:names_fluxerror}}
    \addtolength{\tabcolsep}{-2pt}
    \begin{tabular}{clcc}
        \hline
        Reference ${G}_{\rm mag}$  &  \Gaia{} EDR3  & ${G}_{\rm mag}$ & Median Error \\ 
        \hline  \hline
        12 & 4019458647338779648 & 12.49 & 0.05 \\
	13 & 3348071631670500736 & 13.00 & 0.09 \\
        14 & 866719456827107712  & 14.17 & 0.10 \\
        15 & 958262527212954752  & 14.98 & 0.13 \\
        16 & 39124751182907520   & 16.01 & 0.17 \\
        17 & 1853351441328008192 & 17.02 & 0.28 \\
        18 & 1199686173677816576 & 18.01 & 0.54 \\
        19 & 3705413559233426432 & 19.07 & 0.97 \\
        20 & 1307679071887310592 & 20.58 & 2.61 \\
        \hline
    \end{tabular}
\end{table}

\subsection{Calcium Detection Limits} \label{sec:ca_limits}

In this section, we sought to quantify the sensitivity of the LAMOST spectrograph to the presence of calcium pollution. To this end, we studied the effects of three properties on the instrument's detection efficiency, namely: the white dwarf's effective temperature, its \Gaia{} ${G}_{\rm mag}$-magnitude, and the amount of calcium in its photosphere. To limit the scope of our analysis, we only considered He-envelope white dwarfs with effective temperatures between 10,000~K and 20,000~K, logarithmic surface gravities between 7 and 9 (with $g$ in cgs units of cm·s$^{-2}$), and a variety of photospheric compositions spanning a high-dimensional grid with 11 metal abundances relative to helium (i.e. H, Ca, Mg, Fe, O, Si, Ti, Be, Cr, Mn, and Ni). For calcium specifically, we considered calcium logarithmic abundances (in base 10) from -12 to -7 \cite{Coutu:2019}. We note that our temperature cut at 10,000~K did not allow us to probe the low-temperature regime where absorption lines can become very broad in cool, He-atmosphere white dwarfs due to collisional pressure broadening \citep{Hollands:2017, Blouin:2018b}. However, we expect cool DZs to be readily identifiable at low temperatures with LAMOST spectroscopy.

Upon defining the hyperparameters of our simulation, we used the model atmosphere code described in \citealt{Dufour:2007, Blouin:2018a, Blouin:2018b} and references therein to produce 1,000 high-resolution synthetic spectra (i.e. 55,000 wavelength points between 3,000 and 9,000~\angs). This code employs all the necessary constitutive physics of white dwarfs to predict their absorption features and has been used successfully in the past to model real UV and optical observations of metal-polluted white dwarfs (e.g. \citealt{Kaiser:2021, Klein:2021, Caron:2023, Doyle:2023}). After generating our synthetic models, we convolved them to a resolving power of $R=1,800$ with a Gaussian kernel, and used the \texttt{SpecTres}\footnote{\url{https://spectres.readthedocs.io/en/latest/}.} \texttt{Python} package to downsample them to the wavelength grid of LAMOST observations.

In addition to smoothing and resampling our synthetic spectra, we also injected a representative amount of LAMOST Gaussian noise into the data. To achieve this, we started by identifying all the LAMOST stars with \Gaia{} ${G}_{\rm mag}$-magnitudes within 0.1 units of the integer values \Gaia{} ${G}_{\rm mag}\subset[12,20]$. This magnitude range encompassed all the white dwarf candidates in our catalogue, from the brigthest to the faintest object (${G}_{\rm mag}\approx12.5$ and ${G}_{\rm mag}\approx20.6$, respectively). Then, for a each integer magnitude, we selected all the associated LAMOST spectra and calculated the median of their flux errors between 3,800~\angs{} and 4,000~\angs. Based on these median results, we picked the spectrum with the most representative flux error (see Table \ref{tab:names_fluxerror}) and injected the latter into the 1,000 atmosphere models. \autoref{fig:montreal} shows the effects of downgrading, resampling, and adding LAMOST noise to a synthetic spectrum of a He-rich white dwarf with effective temperature \Teff=11,066~K, surface gravity \logg=7.30 cgs, and calcium abundance \logCaHe=-9.50.

Using our Lorentzian algorithm to identify calcium pollution (see Section \ref{sec:lorentz_classification}), we  generated the detection limits show in \autoref{fig:ca_threshold} for each reference \Gaia{} ${G}_{\rm mag}$-magnitude. As expected, LAMOST performs well for bright and cool white dwarfs, but loses significant sensitivity at higher magnitudes and effective temperatures. This behaviour is expected: on the one hand, fainter objects tend to exhibit noisier spectra, which hampers the detection of calcium at low resolution; on the other hand, hotter white dwarfs have atmospheres that are more opaque, which makes it more difficult to detect low levels of calcium pollution in their observations.

\begin{figure*} 
    \begin{center}
        \includegraphics[width=0.96\linewidth]{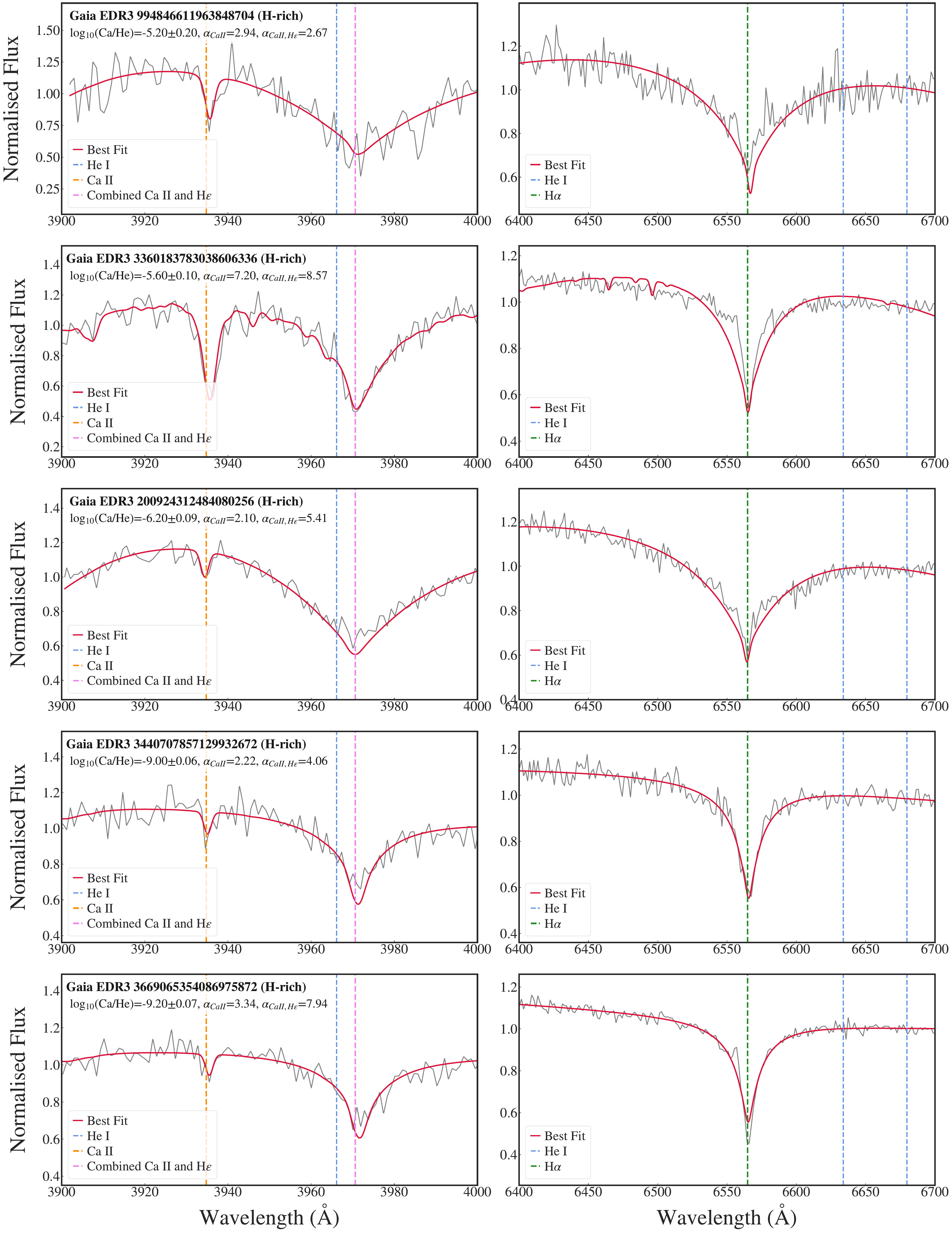}
        \caption{LAMOST spectra (in black) and spectroscopic models (in red) for 5 H-dominated calcium-polluted white dwarfs. The top three panels show the most polluted H-rich stars considered in our atmospheric analysis, while the bottom two panels present our best-fit models for the least two polluted H-rich stars. In each panel, the left and right figures illustrate the Ca II and He I/H-$\alpha$ regions between 3,900-4,000~\angs{} and 6,400-6,700~\angs, respectively.}
        \label{fig:H_fit}
    \end{center}
\end{figure*}

\begin{figure*} 
    \begin{center}
        \includegraphics[width=0.96\linewidth]{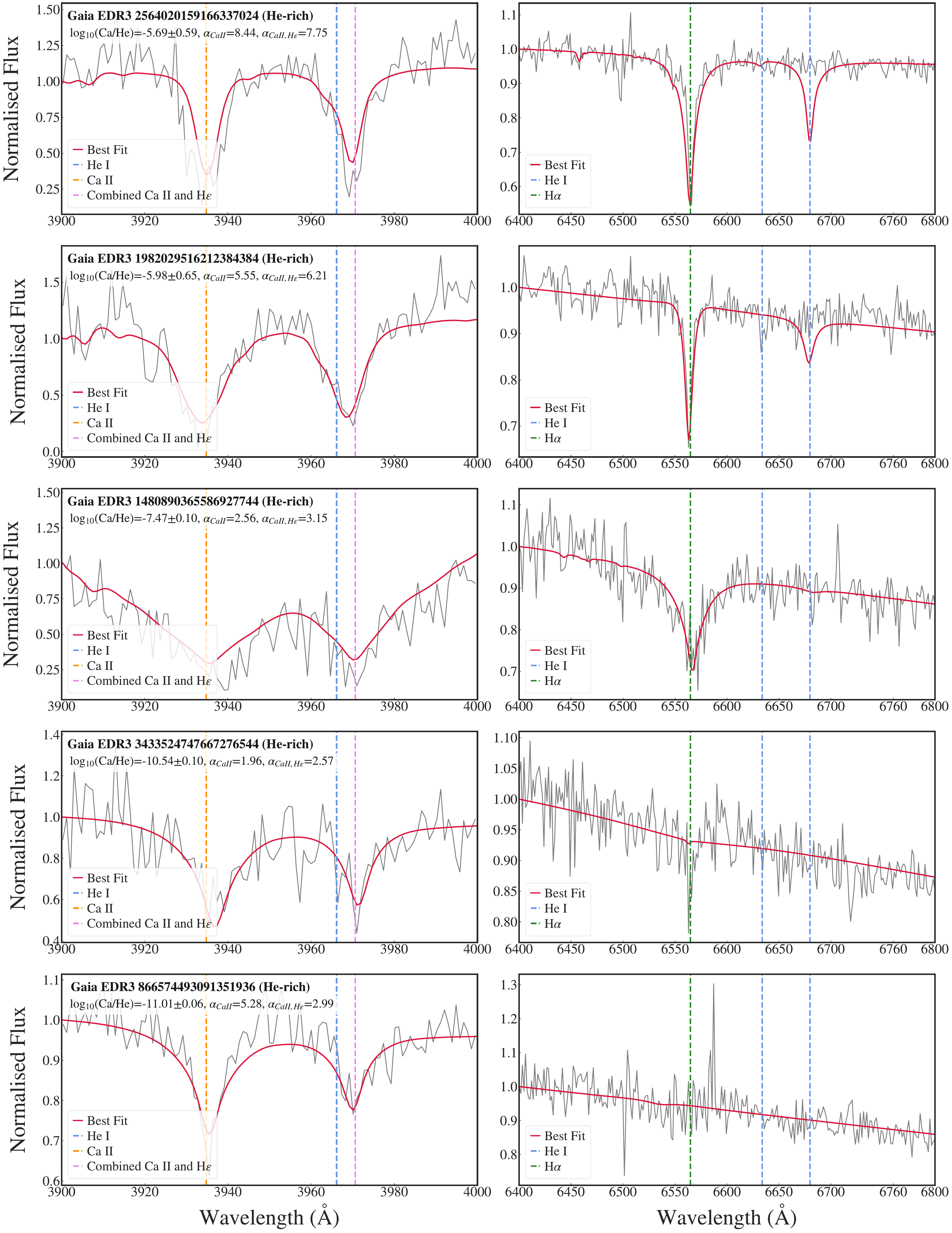}
         \caption{LAMOST spectra (in black) and spectroscopic models (in red) for 5 He-dominated calcium-polluted white dwarfs. The top three panels show the most polluted He-rich stars considered in our atmospheric analysis, while the bottom two panels present our best-fit models for the least two polluted He-rich stars. In each panel, the left and right figures illustrate the Ca II and He I/H-$\alpha$ regions between 3,900-4,000~\angs{} and 6,400-6,700~\angs, respectively.}
        \label{fig:He_fit}
    \end{center}
\end{figure*}

\begin{figure*}
 \centering
        \includegraphics[width=1\textwidth]{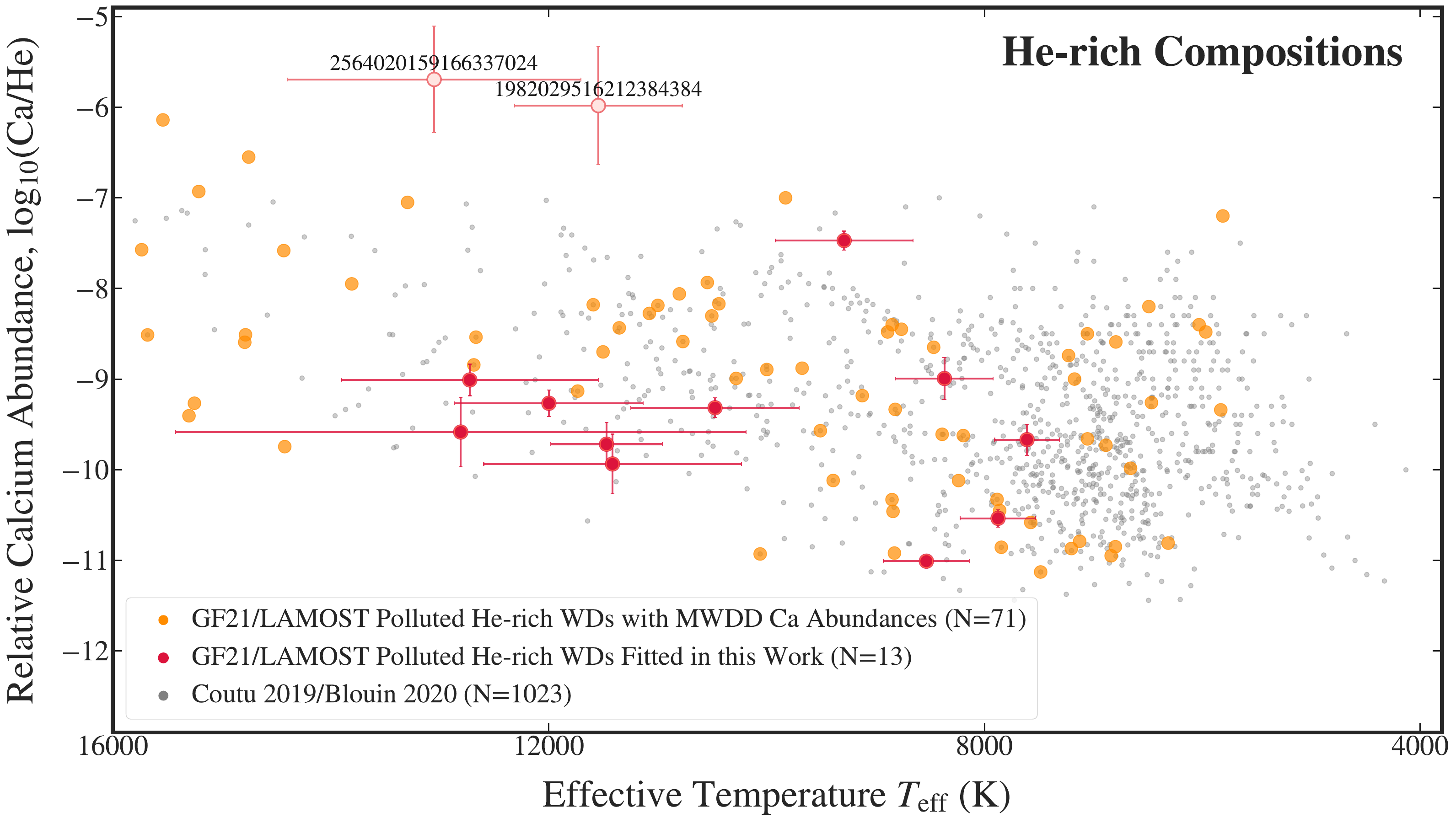}
        \caption{Calcium abundances \logCaHe{} as a function of effective temperature for the 13 He-dominated polluted white dwarfs fitted in this work (in red, see Section \ref{sec:atmospheric_analysis}). We also include 71 GF21/LAMOST white dwarfs with calcium abundances in the MWDD (in orange), as well as 1,023 DBZ/DZ(A) stars from the study of \citealt{Coutu:2019}/\citealt{Blouin:2020} (in grey). As discussed in Section \ref{sec:atmospheric_analysis}, our best-fit calcium abundances for the two objects in the upper left corner of this figure may be overestimated, so we show them in lighter red. We note that we do not include the \logCaHe{} uncertainties of the 71 GF21/LAMOST white dwarfs because these errors are not reported in the MWDD. }
        \label{fig:ca_vs_teff}
\end{figure*}

\subsection{Atmospheric Analysis} \label{sec:atmospheric_analysis}

The last step in our work was to determine the main physical properties and calcium abundances of our final catalogue of polluted white dwarfs. For this purpose, we  only considered those stars that were either unknown to the MWDD, or that were listed in the MWDD but had no existing calcium abundances. Moreover, for those sources with known effective temperatures, we further restricted our sample to \Teff$\leq25,000$~K. This temperature threshold ensured that the origin of the observed calcium pollution was due to the accretion of exoplanetary bodies, rather than  radiative levitation bringing metals to the photosphere. To perform our temperature cut, we used the \Teffmix{} measurements of GF21, and in the absence of the latter, we considered the \Teffmontreal{} values in the MWDD. 

Among the 62 white dwarfs in our polluted sample, 28 satisfied the inclusion criteria for atmospheric analysis. However, we had to remove 5 of them from our study after conducting preliminary fits to their existing photometry and/or LAMOST spectra. More specifically, we excluded 2 stars with an effective temperature greater than \Teff$\geq$25,000~K according to their \panstarrs{} or SDSS photometry, and 3 objects lacking a consistent spectroscopic solution. Among the latter, 2 had a relatively low surface gravity based on our initial photometric fits (\logg{}$\lesssim$7.1), which could be a sign that they are unresolved binaries. Table \ref{tab:discarded_in_ca_fits} lists the \Gaia{} EDR3 names and GF21 P$_{\rm WD}$ probabilities of the 5 discarded objects.

For the 23 polluted white dwarfs suitable for our atmospheric analysis, we started by classifying them as H- or He-rich based on the dominant absorption lines in their spectra and the width of these lines. This allowed us to identify 10 and 13 H- and He-rich stars, respectively. We then assumed a pure hydrogen atmosphere with a trace of metals if hydrogen was their main chemical constituent, or a pure helium atmosphere with metal pollutants if helium was their predominant component. Next, we employed a standard spectro-photometric fitting method \citep[e.g.,][]{Dufour:2007} and the model atmospheres of \citet[and references therein]{Blouin:2018a,Blouin:2018b} to determine their main physical properties and calcium abundances. 

As an initial step, we fitted the synthetic fluxes to the \panstarrs{} photometry (or, if unavailable, to the SDSS photometry) to infer their \Teff{} and solid angle $\pi R^2/D^2$. Note that for this task, we first had to de-redden the photometry from 3D interstellar extinction maps \citep{Lallement:2019,GentileFusillo:2021}. After concluding the photometric fits, we used the \Gaia{} EDR3 parallax of each source to calculate its radius $R$ from the solid angle determined with the photometric fit. From there, we used the evolutionary models of \cite{Bedard:2020} to obtain the mass $M$ and surface gravity \logg{} of the white dwarf. For this last step, we assumed a thick hydrogen envelope $M_{\rm H}/M_{\star} = 10^{-4}$ for objects with an H-rich atmosphere, and a thin hydrogen envelope $M_{\rm H}/M_{\star} = 10^{-10}$ for He-atmosphere objects. Fixing \Teff{} and \logg, we adjusted the calcium abundance of each WD candidate to fit its LAMOST spectrum in the calcium region between 3,800~\angs{} and 4,000~\angs. Fixing Ca/H(e) to that value, we repeated the photometric fit. Using the updated \Teff{} and \logg, we adjusted Ca/H(e) once more, and we repeated the whole procedure until we found no variations in our free parameters and the latter converged to well-defined values. 

In our model atmospheres, we included a wavelength shift to account for possible offsets in the spectral lines caused by wavelength calibration issues, stellar barycentric radial velocity, and gravitational redshift. We also included all the metals from C to Cu, but only adjusted Ca to the LAMOST data. For the rest of the metals, we assumed chondritic abundance ratios relative to calcium \citep{Lodders:2003}. This assumption speeds up the computation of model atmosphere grids and is typically used as a good first-order approximation when the  atmospheric composition of the white dwarf is unknown  (e.g. \citealt{Blouin:2020, Coutu:2019, Dufour:2007}). Observationally, this practice may be justified on two grounds: first, most exoplanetesimals are expected to have chondritic/primordial compositions, and second, the majority of known polluted white dwarfs exhibit abundances consistent with the accretion of chondrite-like material \citep[e.g.][]{Harrison:2021}.

The results of our atmospheric analysis\footnote{As is standard in white dwarf characterisation work, we report our abundance measurements as number abundances.} are presented in Table \ref{tab:fit_results}  and can be downloaded in machine-readable format from our GitHub repository.\footnote{\url{https://github.com/mbadenas/gaialamost}}} In Figures \ref{fig:H_fit} and \ref{fig:He_fit}, we also show a selection of spectroscopic fits for white dwarfs of different photospheric compositions. According to our models, the H-rich white dwarfs with the highest levels of calcium pollution are \Gaia{} EDR3 994846611963848704, \Gaia{} EDR3 3360183783038606336, and \Gaia{} EDR3 200924312484080256 \citep{Jimenez:2018}, with \logCaH{} abundances corresponding to about $-5.20$, $-5.60$, and $-6.20$, respectively. Among the group of He-dominated white dwarfs, the most heavily calcium-polluted objects are \Gaia{} EDR3 2564020159166337024 \citep{Jimenez:2018}, \Gaia{} EDR3 1982029516212384384 \citep{Jimenez:2018}, and \Gaia{} EDR3 1480890365586927744 \citep{Girven:2011}, with \logCaHe{} values ranging from about $-5.70$ to $-7.50$. We note, however, that our best models for \Gaia{} EDR3 2564020159166337024 and \Gaia{} EDR3 1982029516212384384 provide a poor fit to the spectrum around the helium I line at 6,690~\angs{} (see \autoref{fig:He_fit}). This observation, combined with the relatively small surface gravity of these two stars, may be an indication that they are unresolved binaries. Therefore, we treat their calcium abundances with caution. For the remaining polluted white dwarfs, our spectroscopic models are consistent with the observed LAMOST spectra, hence offering a first glimpse into the fundamental properties of their atmospheres.

\begin{landscape}
\begin{table}
    \caption{Main properties of the 23 polluted white dwarfs considered in our atmospheric analysis. Among this sample, there are 19 stars with previously unknown traces of calcium pollution (in blue). The full table contains additional columns and is available in machine-readable form in the online supplementary material. The notation $\log$(Ca/Y) stands for the logarithm (in base 10) of the number abundance ratio between calcium and $Y$. The superscript letter provides the discovery paper of the star. The  asterisk symbol (${*}$) denotes potential unresolved binaries (see discussion in Section \ref{sec:atmospheric_analysis}), while the symbol ``$\ddagger\ddagger$'' is reserved for intermediate-temperature (8,000$\leq$\Teff$\leq$25,000~K), moderately bright ($G_{\rm mag}\leq17.5$) polluted sources suitable for follow-up spectroscopy (see Section \ref{sec:observational_priorities}). Lastly, the \textsc{Mg?} column shows whether the star has possible signs of magnesium pollution, with ``Y,'' ``N,'' and ``A'' standing for ``Yes,'' ``No,'' and ``Ambiguous,'' respectively (see Section \ref{sec:other_elements}).}
    \label{tab:fit_results}
    \addtolength{\tabcolsep}{-5.4pt}    
    \renewcommand{\arraystretch}{1.2}
    \begin{tabular}{|c|lc|cc|cc|ccc|cc|c|}
    \hline
      & \Gaia{} EDR3  & $G_{\rm mag}$ & $T_{\rm eff, MWDD}$  & $T_{\rm eff, This Work}$    & $\log(g)_{\rm MWDD}$ & $\log(g)_{\rm This Work}$  & $\log(\rm{H/He})_{\rm This Work}$ & $\log(\rm{Ca/H})_{\rm This Work}$ &  $\log(\rm{Ca/He})_{\rm This Work}$  &  Mass$_{\rm MWDD}$ & Mass$_{\rm This Work}$ & Mg? \\
    \hdashline[0.2pt/1pt]
     \small{No. in Tab. \ref{tab:props_polluted}} &  &  & [K]  & [K]    & [cgs] & [cgs]   &  &  &  &  [\Msun]  & [\Msun] & \\
    \hline \hline
    \multicolumn{13}{|c|}{\it Hydrogen-Rich White Dwarfs}       \\
    \hline
    62 & \blue{994846611963848704}$^{\tiny{a}}$                          & 17.71 & -            & 12449.77$\pm$168.94 & -             & 7.98$^{+0.03}_{-0.03}$ & -              & -5.20$\pm$0.20 & -               & -             & 0.59$^{+0.02}_{-0.02}$ & N \\  
    29 & \blue{3360183783038606336}$^{\tiny{a}}$$^{\ddagger\ddagger}$    & 16.60 & 9516         & 9416.70$\pm$98.67   & 7.98          & 7.99$^{+0.02}_{-0.02}$ & -              & -5.60$\pm$0.10 & -               & 0.58          & 0.59$^{+0.01}_{-0.01}$ & N \\  
    12 & \blue{200924312484080256}$^{\tiny{a}}$                          & 17.52 & -            & 12665.43$\pm$215.35 & -             & 7.73$^{+0.03}_{-0.03}$ & -              & -6.20$\pm$0.09 & -               & -             & 0.47$^{+0.01}_{-0.01}$ & N \\  
    45 & \blue{642549544391197440}$^{\tiny{b}}$$^{\ddagger\ddagger}$     & 16.47 & 10919$\pm$66 & 10875.18$\pm$114.07 & 7.97          & 7.99$^{+0.02}_{-0.02}$ & -              & -6.62$\pm$0.06 & -               & 0.59$\pm$0.01 & 0.60$^{+0.01}_{-0.01}$ & N \\  
    15 & \blue{2503322616188632320}$^{\tiny{f}}$         & 20.19 & -    & 5643.00$\pm$655.0   & -     & 7.81$^{+0.49}_{-0.49}$  &    -  & -7.30$\pm$0.60 &  -      & -             & 0.50$^{+0.29}_{-0.19}$ & Y  \\ 
    35 & \blue{367949367212923392}$^{\tiny{c}}$$^{\ddagger\ddagger}$     & 17.13 & 9830$\pm$53  & 9597.09$\pm$69.03   & 7.98$\pm$0.02 & 7.95$^{+0.02}_{-0.02}$ & -              & -7.94$\pm$0.06 & -               & 0.59$\pm$0.01 & 0.57$^{+0.01}_{-0.01}$ & N \\  
    10 & \blue{1565012935076040832}$^{\tiny{b}}$                         & 16.72 & 7583$\pm$20  & 7462.04$\pm$28.75   & 8.12$\pm$0.01 & 8.09$^{+0.01}_{-0.01}$ & -              & -8.03$\pm$0.05 & -               & 0.66$\pm$0.00 & 0.65$^{+0.01}_{-0.01}$ & N \\  
    20 & \blue{2759588063311504768}$^{\tiny{b}}$                         & 17.27 & 6836$\pm$39  & 6756.32$\pm$23.51   & 7.98$\pm$0.01 & 7.97$^{+0.01}_{-0.01}$ & -              & -8.72$\pm$0.05 & -               & 0.58$\pm$0.01 & 0.58$^{+0.01}_{-0.01}$ & N \\  
    32 & \blue{3440707857129932672}$^{\tiny{b}}$                         & 17.19 & 7982         & 7898.84$\pm$67.18   & 7.96          & 7.96$^{+0.02}_{-0.02}$ & -              & -9.00$\pm$0.06 & -               & 0.57          & 0.57$^{+0.01}_{-0.01}$ & N \\  
    34 & \blue{3669065354086975872}$^{\tiny{d}}$                         & 17.01 & 7980$\pm$37  & 7728.46$\pm$58.42   & 8.08$\pm$0.01 & 8.03$^{+0.02}_{-0.02}$ & -              & -9.20$\pm$0.07 & -               & 0.64$\pm$0.01 & 0.61$^{+0.01}_{-0.01}$ & N \\  
    \hline
   \multicolumn{13}{|c|}{\it Helium-Rich White Dwarfs}       \\
    \hline
    16 & \blue{2564020159166337024}$^{\tiny{a}}$$^{*}$                  & 16.82 & -            & 13050$\pm$1346      & -             & 7.68$^{+0.08}_{-0.07}$ & -3.82$\pm$0.16 & -              & -5.69$\pm$0.59  & -             & 0.42$^{0.04}_{-0.03}$ & Y \\  
    11 & \blue{1982029516212384384}$^{\tiny{a}}$$^{*}$                  & 17.18 & -            & 11543$\pm$770       & -             & 7.50$^{+0.12}_{-0.11}$ & -5.33$\pm$0.12 & -              & -5.98$\pm$0.65  & -             & 0.34$^{0.05}_{-0.04}$ & Y \\  
    8  & \blue{1480890365586927744}$^{\tiny{b}}$                        & 17.76 & -            & 9287$\pm$633        & -             & 8.16$^{+0.11}_{-0.11}$ & -4.34$\pm$0.16 & -              & -7.47$\pm$0.10  & -             & 0.68$^{+0.07}_{-0.07}$ & Y \\  
    42 & \blue{4502205519647039616}$^{\tiny{a}}$                        & 17.75 & 9173         & 8367$\pm$447        & 8.27          & 7.99$^{+0.10}_{-0.09}$ & -3.93$\pm$0.21 & -              & -8.99$\pm$0.23  & 0.76          & 0.57$^{+0.06}_{-0.05}$ & N \\  
    61 & \blue{972551088836290816}$^{\tiny{a}}$$^{\ddagger\ddagger}$     & 17.31 & -            & 12723$\pm$1179      & -             & 7.83$^{+0.16}_{-0.15}$ & -4.65$\pm$0.11 & -              & -9.01$\pm$0.17  & -             & 0.49$^{+0.09}_{-0.07}$ & A \\  
    25 & 322324719900568448$^{\tiny{a}}$$^{\ddagger\ddagger}$            & 17.32 & -            & 11997$\pm$864       & -             & 7.95$^{+0.11}_{-0.10}$ & -5.51$\pm$0.20 & -              & -9.27$\pm$0.15  & -             & 0.56$^{+0.06}_{-0.06}$ & A \\  
    54 & 839750257663742464$^{\tiny{e}}$$^{\ddagger\ddagger}$            & 16.75 & 10754        & 10472$\pm$773       & 8.2           & 8.07$^{+0.12}_{-0.12}$ & -5.39$\pm$0.13 & -              & -9.32$\pm$0.11  & 0.73          & 0.62$^{+0.08}_{-0.07}$ & N \\  
    28 & \blue{3351139990665573120}$^{\tiny{a}}$                        & 17.65 & -            & 12806$\pm$2617      & -             & 7.65$^{+0.50}_{-0.38}$ & -3.95$\pm$0.60 & -              & -9.58$\pm$0.38  & -             & 0.41$^{+0.27}_{-0.13}$ & N \\  
    7  & \blue{1411668854417599104}$^{\tiny{d}}$                        & 18.89 & 7930         & 7609$\pm$297        & 8.05          & 7.90$^{+0.09}_{-0.09}$ & -1.5$\pm$6.75  & -              & -9.67$\pm$0.17  & 0.61          & 0.52$^{+0.05}_{-0.05}$ & N \\  
    24 & \blue{321093335597030400}$^{\tiny{f}}$                         & 18.70 & -            & 11468$\pm$511       & -             & 7.92$^{+0.12}_{-0.12}$ & -5.01$\pm$0.48 & -              & -9.72$\pm$0.24  & -             & 0.54$^{+0.07}_{-0.06}$ & N \\  
    26 & 3253852143823041792$^{\tiny{a}}$                               & 17.17 & -            & 11413$\pm$1182      & -             & 7.88$^{+0.19}_{-0.18}$ & -5.31$\pm$0.18 & -              & -9.94$\pm$0.33  & -             & 0.52$^{+0.11}_{-0.09}$ & N \\  
    31 & \blue{3433524747667276544}$^{\tiny{a}}$                        & 17.38 & 8672         & 7875$\pm$346        & 8.35          & 8.09$^{+0.08}_{-0.08}$ & -4.0$\pm$0.26  & -              & -10.54$\pm$0.10 & 0.82          & 0.63$^{+0.05}_{-0.05}$ & N \\  
    55 & 866574493091351936$^{\tiny{f}}$                                & 17.45 & 8787$\pm$64  & 8533$\pm$395        & 8.06$\pm$0.02 & 8.08$^{+0.09}_{-0.09}$ & -5.52$\pm$1.36 & -              & -11.01$\pm$0.06 & 0.61$\pm$0.01 & 0.63$^{+0.06}_{-0.05}$ & N \\  
     \hline
    \end{tabular}
    \begin{quote}
    \hspace{0.4cm} White Dwarf Discovery Reference: [a]: \citealt{Jimenez:2018}, [b]: \citealt{Girven:2011}, [c]: \citealt{Guo:2022}, [d]: \citealt{Eisenstein:2006}, [e]: \citealt{Abazajian:2009}, [f]: \citealt{GentileFusillo:2015}.
    \end{quote}
\end{table}
\end{landscape}
\addtolength{\tabcolsep}{1pt}

In \autoref{fig:ca_vs_teff}, we also present our calcium abundance measurements as a function of effective temperature for the 13 He-rich polluted white dwarfs considered in our atmospheric analysis. For comparison purposes, we also include the sample of GF21/LAMOST white dwarfs with He-compositions and known calcium abundances (N=71), as well as the results of \citealt{Coutu:2019}, who performed a comprehensive study of $1,023$ DBZ/DZ(A) He-rich white dwarfs using model atmosphere codes similar to those employed in this work. When available, however, we replaced their calcium estimates with those of \citealt{Blouin:2020}. In both cases, the authors used the same atmosphere code, but while the former only adjusted Ca in their model (assuming chondritic abundance ratios for other metals), the latter performed a simultaneous fit of the Ca, Fe, and Mg lines, which can result in non-negligible changes to the atmospheric parameters. 

As observed in \autoref{fig:ca_vs_teff}, our distribution of He-dominated polluted white dwarfs is generally consistent with that of \citealt{Coutu:2019}/\citealt{Blouin:2020}. However, there are 2 stars with calcium abundances significantly larger than the average value of \logCaHe$\approx$-8.09 at around 12,000~K, namely: \Gaia{} EDR3 2564020159166337024 and \Gaia{} EDR3 1982029516212384384. As mentioned above, these objects may be unresolved white dwarf binaries with unreliable calcium abundance measurements. Another important characteristic of \autoref{fig:ca_vs_teff} is the scarcity of hot, He-rich white dwarfs with low levels of calcium pollution. This pattern is not caused by a real relationship between the amount of accreted material and the temperature of the star, but by an observational bias arising from the detection limit of calcium as a function of \Teff{} in He-dominated objects \citep{Hollands:2017, Coutu:2019, Blouin:2021}. At lower temperatures, it is easier to detect lower levels of calcium pollution because the photospheres are more transparent; this is evidenced by the presence of several GF21/LAMOST polluted stars with \logCaHe$\leq$-9. Finally, we note that the 13 He-dominated white dwarfs in \autoref{fig:ca_vs_teff} have a mean calcium abundance (\logCaHe$\approx$-9.50) that is only slightly lower than that of the 71 He-rich GF21/LAMOST stars with known calcium abundances in the MWDD (\logCaHe$\approx$-9.03). However,  \autoref{fig:ca_vs_teff} does not reveal any clear pattern in our abundance measurements compared to those reported by the MWDD.

\subsection{Observational Priorities} \label{sec:observational_priorities}

As demonstrated in this work, the low-resolution spectra of LAMOST do not seem to reveal clear traces of heavy elements beyond calcium. Nevertheless, despite this limitation, LAMOST constitutes a valuable tool to efficiently screen a large number of white dwarfs and identify those stars with spectral features warranting further investigation. In this section, we use the results of our atmospheric analysis (see Section \ref{sec:atmospheric_analysis}) to select the most promising polluted white dwarfs for detailed characterisation work with high-resolution spectroscopy. For this search, we imposed two criteria on the stars shown in Table \ref{tab:fit_results}:

\begin{itemize}[noitemsep,topsep=1pt,leftmargin=14pt]
    \item First, we required a $\Gaia{}$ magnitude lower than $G_{\rm mag}\leq$17.5. At higher magnitudes (i.e. fainter objects), longer exposure times are needed to obtain a good S/N and detect calcium pollution, hence hindering the viability and success of an observing proposal. 
    \item Second, we looked for stars with effective temperatures between 8,000~K and 25,000~K. At temperatures lower than \Teff$\lesssim$8,000~K, fewer metallic species are detectable since the thermal energy is insufficient to populate many of the atomic levels responsible for transitions in the optical. Although several cool white dwarfs have been found with different metals in their spectra, the majority of well-characterised polluted white dwarfs are hotter than 8,000~K. Therefore, given our interest in detecting \textit{multiple} heavy elements for planetary system characterisation, we decided to omit the coolest objects in our search. Regarding our temperature cut at 25,000~K, the selection criteria imposed in Section \ref{sec:atmospheric_analysis} ensured that all the stars in Table \ref{tab:fit_results} were cooler than this threshold.  
\end{itemize}

Excluding the two stars with excessively high calcium abundances in the upper left corner of \autoref{fig:ca_vs_teff}, and applying the two aforementioned criteria to Table \ref{tab:fit_results}, we identified the top three H- and He-dominated polluted white dwarfs for high-resolution spectroscopy; these sources are flagged with the symbol ``$\ddagger\ddagger$'' in Table \ref{tab:fit_results}. Among them, there are 4 white dwarfs with previously unknown signs of calcium pollution (i.e. \Gaia{} EDR3 3360183783038606336 \citep{Jimenez:2018}, \Gaia{} EDR3 642549544391197440 \citep{Girven:2011}, \Gaia{} EDR3 367949367212923392 \citep{Guo:2015}, and \Gaia{} EDR3 972551088836290816 \citep{Jimenez:2018}, as well as 2 calcium-polluted white dwarfs with potential traces of magnesium (\Gaia{} EDR3 972551088836290816 and \Gaia{} EDR3 322324719900568448 \citep{Jimenez:2018}. Based on the cross-matching procedure with the MAST portal described in Section \ref{sec:properties_catalogue}, none of the selected targets have HST observations, so they are all interesting candidates for future UV campaigns.

\section{Discussion and Conclusion}  \label{sec:discussion_and_conclusions}

The analysis presented in this paper is the first attempt at using the low-resolution LAMOST spectroscopic survey for the search and characterisation of white dwarf metal pollution.

To begin with, we cross-matched the \Gaia{} EDR3 database of \citealt{GentileFusillo:2021} with the LAMOST LRS DR9v1 archive, obtaining a catalogue of 4,324 unique white dwarf candidates (Section  \ref{sec:methodology}). Next, we used a Lorentzian model to assess the presence of calcium pollution in their spectra between 3,800\angs{} and 4,000~\angs{} (Section \ref{sec:lorentz_classification}). From this quantitative analysis, we generated a preliminary sample 76 polluted white dwarf candidates (Section \ref{sec:pollution_threshold}), which we then vetted through a careful visual examination of their LAMOST spectra and multiple literature cross-checks (Section \ref{sec:literature_validation}). Our final polluted sample consists of 62 white dwarfs (1.43$\%$), of which 30 were not previously known to exhibit calcium pollution (Section \ref{sec:final_polluted_sample}).

For our final sample of polluted systems, we also estimated preliminary atmospheric parameters for 23 white dwarfs with no existing calcium measurements in the MWDD and effective temperatures in the range 6,750~K$\leq$\Teff$\leq$13,000~K. For these stars, our paper represents an initial step towards determining their full chemical compositions with future high-resolution spectroscopy. This follow-up work will be crucial to increase the number of well-characterised polluted white dwarfs and move towards population-wide studies of extrasolar material. In Section \ref{sec:observational_priorities}, we list several promising targets for future observations. 

To put in perspective the number of polluted \Gaia{} EDR3 white dwarfs in LAMOST DR9v1, it is useful to consider the case of SDSS. This imaging and spectroscopic survey was specifically designed to observe distant objects in the northern Galactic gap, such as galaxies and quasars \citep{York:2000}. However, thanks to its optimal performance in the blue, it has also been extensively used to spectroscopically confirm and generate catalogues of white dwarfs. For instance, \citealt{Kleinman:2013} and \citealt{Kepler:2021} identified 1,316 unique DZ sources with SDSS spectra across multiple data releases (DR7-DR16), or 3.6$\%$ of the total number of objects that they classified as white dwarfs (30,086 DAs, 2,390 DCs, 2,160 DBs, 1,316 DZs, 572 DQs, 137 DOs, 4 DS). Based on our work, the number of polluted white dwarfs detected by LAMOST is somewhat lower than in SDSS, likely due to different observing strategies.

Over the past two decades, SDSS has demonstrated that low-resolution spectroscopic surveys can play a key role in the study of polluted white dwarfs. As we have also shown with LAMOST, these surveys are particularly useful to obtain preliminary estimates of WD elemental abundances and inform higher-resolution characterisation campaigns. With the advent of wide-field, multi-epoch spectroscopic surveys such as DESI \citep{DESI} and WEAVE \citep{WEAVE} in the north, 4MOST in the south \citep{4MOST}, and SDSS V \citep{SDSS_V} across the entire sky, LAMOST has the potential to become an important tool in the detection of WD heavy element  pollution. By building synergies with these  next-generation spectrographs, LAMOST will contribute to expanding the population of polluted white dwarfs, therefore helping to advance our statistical understanding of their accreted material and of extrasolar geochemistry more broadly. 

\section*{Acknowledgements}
We thank the anonymous referees for their insightful feedback and valuable suggestions, which improved the quality of the original manuscript.

MBA is supported by the MIT Department of the Earth, Atmospheric, and Planetary Sciences, NASA grants 80NSSC22K1067 and 80NSSC22K0848, and the MIT William Asbjornsen Albert Memorial Fellowship. SB is a Banting Postdoctoral Fellow and a CITA National Fellow, supported by the Natural Sciences and Engineering Research Council of Canada (NSERC). 

This work has made use of data from the European Space Agency (ESA) mission \Gaia{} (\url{https://www.cosmos.esa.int/gaia}), processed by the \Gaia{} Data Processing and Analysis Consortium (DPAC,
\url{https://www.cosmos.esa.int/web/gaia/dpac/consortium}). Funding for the DPAC has been provided by national institutions, in particular the institutions participating in the \Gaia{} Multilateral Agreement. 

This work has also made use of low-resolution spectra acquired by the Guoshoujing Telescope (the Large Sky Area Multi-Object Fiber Spectroscopic Telescope LAMOST), a National Major Scientific Project built by the Chinese Academy of Sciences. Funding for the project has been provided by the National Development and Reform Commission. LAMOST is operated and managed by the National Astronomical Observatories, Chinese Academy of Sciences.

Moreover, this work has made use of data from the Sloan Digital Sky Survey (SDSS), which is funded by the Alfred P. Sloan Foundation, the Heising-Simons Foundation, the National Science Foundation, and the Participating Institutions. SDSS acknowledges support and resources from the Center for High-Performance Computing at the University of Utah. The SDSS web site is \url{www.sdss.org}. SDSS is managed by the Astrophysical Research Consortium for the Participating Institutions of the SDSS Collaboration, including the Carnegie Institution for Science, Chilean National Time Allocation Committee (CNTAC) ratified researchers, the Gotham Participation Group, Harvard University, Heidelberg University, The Johns Hopkins University, L’Ecole polytechnique Fédérale de Lausanne (EPFL), Leibniz-Institut für Astrophysik Potsdam (AIP), Max-Planck-Institut für Astronomie (MPIA Heidelberg), Max-Planck-Institut für Extraterrestrische Physik (MPE), Nanjing University, National Astronomical Observatories of China (NAOC), New Mexico State University, The Ohio State University, Pennsylvania State University, Smithsonian Astrophysical Observatory, Space Telescope Science Institute (STScI), the Stellar Astrophysics Participation Group, Universidad Nacional Autónoma de Mexico, University of Arizona, University of Colorado Boulder, University of Illinois at Urbana-Champaign, University of Toronto, University of Utah, University of Virginia, Yale University, and Yunnan University.

This work has made use of data from the Pan-STARRS1 Surveys (PS1) and the PS1 public science archive. The PS1 survey has been made possible through contributions by the Institute for Astronomy, the University of Hawaii, the Pan-STARRS Project Office, the Max-Planck Society and its participating institutes, the Max Planck Institute for Astronomy, Heidelberg and the Max Planck Institute for Extraterrestrial Physics, Garching, The Johns Hopkins University, Durham University, the University of Edinburgh, the Queen's University Belfast, the Harvard-Smithsonian Center for Astrophysics, the Las Cumbres Observatory Global Telescope Network Incorporated, the National Central University of Taiwan, the Space Telescope Science Institute, the National Aeronautics and Space Administration under Grant No. NNX08AR22G issued through the Planetary Science Division of the NASA Science Mission Directorate, the National Science Foundation Grant No. AST-1238877, the University of Maryland, Eotvos Lorand University (ELTE), the Los Alamos National Laboratory, and the Gordon and Betty Moore Foundation.

Finally, this work has made use of public data from the Montreal White Dwarf Database, the SIMBAD database and VizieR catalogue (both operated at CDS, Strasbourg, France), the NASA/ESA Hubble Space Telescope (HST) and the Transiting Exoplanet Survey Satellite (TESS) missions. Some of this data were obtained from the Multimission Archive at the Space Telescope Science Institute (MAST). STScI is operated by the Association of Universities for Research in Astronomy, Inc., under NASA contract NAS5-26555. Support for MAST for non-HST data is provided by the NASA Office of Space Science via grant NAG5-7584 and by other grants and contracts.

This work has employed the following open-source software packages: \texttt{Python} (\citealt{Python}), \texttt{numpy} (\citealt{numpy}), \texttt{scipy} (\citealt{scipy}), \linebreak 
\texttt{matplotlib} (\citealt{matplotlib}), \texttt{astropy} (\citealt{astropy:2018}), and \texttt{pandas} (\citealt{pandas:2010}).

\section*{Data Availability} 

Our GF21/LAMOST catalogue of white dwarf candidates (Section \ref{sec:methodology_creation_catalog}) and the results of our atmospheric analysis (Section \ref{sec:atmospheric_analysis}) can be downloaded in machine-readable format from our GitHub repository.\footnote{\url{https://github.com/mbadenas/gaialamost}} The names of these files are ``\texttt{ba24\_unique\_catalog.csv}'' and ``\texttt{ba24\_atmospheric\_analysis}.csv}'', respectively.


\bibliographystyle{mnras}
\bibliography{ref} 

\appendix \label{sec:Appendix}

\section{ }
\begin{table*}
    \label{tab:catalog_example}
    \caption{A selection of 10 rows from our GF21/LAMOST catalogue. The full table, which contains 228 columns, is part of the online supplementary material and can be downloaded from our GitHub repository (see Section ``Data Availability''). The \textsc{RA\_ICRS} and \textsc{DE\_ICRS} columns are taken from GF21 and are given relative to ICRS J2016. The \textsc{obsID} keyword is the LAMOST identifier of the white dwarf candidate. The \textsc{$\alpha_{\mathrm{H8}}$},  \textsc{$\alpha_{\mathrm{CaII}}$}, and \textsc{$\alpha_{\mathrm{CaII,H\epsilon}}$} columns provide, respectively, the statistical significances of the H8, Ca II, and blended Ca II/H$\epsilon$ lines as calculated in Section \ref{sec:lorentz_classification}. Those stars with calcium pollution are flagged with a ``Y'' in column \textsc{CaPollution}. The \textsc{SpectralClass} column  shows our own spectral classification, with the order of the letters reflecting the strength of absorption lines observed in the LAMOST spectrum. Finally, the \textsc{mg\_detection} column indicates the possibility of magnesium pollution in the atmosphere of the white dwarf (``Y'': Yes, ``N'': No, ``A'': Ambiguous.)  } 
    \addtolength{\tabcolsep}{-5.8pt}  
    \begin{tabular}{|lccccccccccccc|}
     \hline  
    $\dots$ &                     GaiaEDR3 &                   $\dots$ &       RA\_ICRS &       DE\_ICRS &    obsID   &         $\dots$ &       $\alpha_{\mathrm{H8}}$ &      $\alpha_{\mathrm{CaII}}$ &       $\alpha_{\mathrm{CaII,H\epsilon}}$ &    CaPollution & SpectralClass & mg\_detection &       $\dots$ \\ 
        \hdashline[0.2pt/1pt] 
        &                     &        &                   [deg]   &       [deg]    &       &        &    &       &         &       &       &                      &      \\                       
        \hline                         
        \hline                
        $\dots$               &        &                   $\dots$ &       &        $\dots$ &        &    $\dots$ &         &       $       \dots$                 &      &                        $\dots$ &                                  &    $\dots$     & \\        
        $\dots$               &        2560009007603950720 &       $\dots$ &        23.36   &        4.02 &       354813204 &       $\dots$ &                      0.07   &                        0.32    &                                  0.04 &           N &         -      & -             &       $\dots$ \\ 
        $\dots$               &        2560454408597246592 &       $\dots$ &        26.21   &        3.40 &       398114122 &       $\dots$ &                      0.30   &                        0.50    &                                  0.36 &           N &         -      & -             &       $\dots$ \\ 
        $\dots$               &        2560852397446990336 &       $\dots$ &        25.32   &        4.59 &       371205168 &       $\dots$ &                      0.78   &                        0.51    &                                  0.59 &           N &         -      & -             &       $\dots$ \\ 
        $\dots$               &        2561185858707723520 &       $\dots$ &        26.58   &        4.73 &       371208148 &       $\dots$ &                      0.832& 0.004                    &       0.46                               &    N           & -         &      - &             $\dots$ \\      
        $\dots$               &        2564020159166337024 &       $\dots$ &        18.04   &        4.92 &       472002174 &       $\dots$ &                      4.05   &                        8.44    &                                  7.75 &           Y &         DAZ    & Y            &       $\dots$ \\ 
        $\dots$               &        2564945432560219008 &       $\dots$ &        23.10   &        5.49 &       354812027 &       $\dots$ &                      1.20   &                        5.36    &                                  4.57 &           Y &         DAZ    & N             &       $\dots$ \\ 
        $\dots$               &        2565499002305389568 &       $\dots$ &        23.97   &        7.05 &       388406035 &       $\dots$ &                      1.24   &                        0.05    &                                  1.08 &           N &         -      & -             &       $\dots$ \\ 
        $\dots$               &        2566210351968389504 &       $\dots$ &        20.54   &        6.74 &       388410068 &       $\dots$ &                      0.55   &                        0.64    &                                  0.91 &           N &         -      & -             &       $\dots$ \\ 
        $\dots$               &        2567791793287585536 &       $\dots$ &        30.55   &        6.94 &       354005104 &       $\dots$ &                      0.01   &                        0.89    &                                  0.46 &           N &         -      & -             &       $\dots$ \\ 
        $\dots$               &        2567899579786157824 &       $\dots$ &        28.86   &        6.71 &       473507018 &       $\dots$ &                      0.63   &                        0.02    &                                  0.73 &           N &         -      & -             &       $\dots$ \\ 
        $\dots$               &        &                   $\dots$ &       &        $\dots$ &        &    $\dots$ &         &       $       \dots$                 &      &                        $\dots$ &                                  &    $\dots$     & \\  
     \hline
    \end{tabular}
    \addtolength{\tabcolsep}{1pt}
\end{table*}

\begin{table*}
    \centering
    \caption{Our initial GF21/LAMOST catalogue of white dwarf candidates consists of 5,556 non-unique stars (see Section \ref{sec:methodology_creation_catalog}). According to the LAMOST 1D Pipeline, 5,092 of them belong to the ``\textsc{WD}'' subclass  (see \textsc{subclass} column in our catalogue). The remaining 464 objects were classified by the 1D Pipeline with the labels shown below.}
    \begin{tabular}{|lccc|}
        \hline
        Type & LAMOST Label & Number & Percentage \\
        \hline
        \hline  
        White dwarf         & \texttt{WD}                 & 5,092 & 91.6\% \\ 
        A-type star         & \texttt{A0}-\texttt{A9}  & 166   & 3.0\% \\
        F-type star         & \texttt{F0}, \texttt{F2}, \texttt{F3}, \texttt{F5-F9}  & 84    & 1.5\% \\
        B-type star         & \texttt{B6}, \texttt{B9}   & 50    & 0.9\% \\
        K-type star         & \texttt{K0}-\texttt{K1} \texttt{K3}-\texttt{K5}, \texttt{K7}    & 40    & 0.7\% \\
        dM/dg-type star     & \texttt{dM0}-\texttt{dM8}, \texttt{gM6} & 31    & 0.6\% \\
        G-type star         &  \texttt{G0}-\texttt{G3}, \texttt{G5}, \texttt{G7}, \texttt{G9}   & 29    & 0.5\% \\
        Non-Classified      &  \texttt{Non}   & 29    & 0.5\% \\
        Cataclysmic Variable  &  \texttt{CV}     & 17 & 0.3\% \\
        O-type star         & \texttt{O}, \texttt{OB}     & 11    & 0.2\% \\
        Double system       &  \texttt{DoubleStar}   & 5     & 0.1\% \\
        Carbon star         &  \texttt{Carbon}  & 1     & $\leq$1\% \\
        EM star             &  \texttt{EM}   & 1     & $\leq$1\% \\
        \hline
        Total               &   -  & 5,556 & 100\% \\
        \hline
    \end{tabular}
    \label{tab:non_wds}
\end{table*}

\begin{table*}
    \centering
    \caption{Main properties of the 14 objects excluded from our final catalogue of 62 calcium-polluted white dwarfs (see Section \ref{sec:literature_validation}). The table provides their LAMOST \textsc{obsID} identifiers, GF21 probabilities (\textsc{$P_{\rm{WD}}$}), celestial coordinates \textsc{RA\_ICRS} and \textsc{DE\_ICRS} (in ICRS J2016),  \textsc{\Teff} and \textsc{\logg} parameters from the MWDD (as of January 2023) and their spectral type reported in the MWDD (when available). The \textsc{Discussion} column references the Section where the object was excluded from the final sample of polluted white dwarfs.}
    \label{tab:discarded_from_catalog}
    \addtolength{\tabcolsep}{-3.5pt}  
    \begin{tabular}{|lcccccccccccc|}
    \hline
     Gaia                  EDR3 &                   obsID &         $P_{\rm{WD}}$ &     RA\_ICRS &      DE\_ICRS &     \Teff{} &                    \logg{} &             $G_{\rm mag}$ &  $\alpha_{\mathrm{H8}}$ & $\alpha_{\mathrm{CaII}}$ & $\alpha_{\mathrm{CaII,H\epsilon}}$ & SpC$_{\rm MWDD}$ &    Discussion \\ 
     \hdashline[0.2pt/1pt] 
                         &    &                   &     [deg]     &             [deg] &        [K]    &        [cgs] &       &                    &       &             &       &     \\ 
     \hline                
     \hline                
     1289860214647954816 &     241503093 &             0.53  &        225.67 &        33.57 &       -                    &       -             &       17.26 &  7.37                   & 9.2                      & 12.83                              & -         &      Sec. 4.3.1\\       
     710040763560788608  &     483108130 &             0.67  &        131.36 &        32.41 &       -                    &       -             &       18.05 &  2.44                   & 6.52                     & 16.90                              & -         &      Sec. 4.3.1\\       
     880821067114616832  &     402906184 &             0.61  &        117.25 &        31.42 &       -                    &       -             &       18.37 &  2.17                   & 2.85                     & 3.74                               & -         &      Sec. 4.3.1\\    
     1598042809833738496 &     906604172 &             0.84  &        233.55 &        54.56 &       34800$\pm$700        &       5.64$\pm$0.09 &       17.12 &  11.74                  & 3.24                     & 16.75                              & sdOB      &      Sec. 4.3.2\\       
     732880265768565248  &     500401142 &             0.90  &        167.59 &        30.79 &       8103                 &       7.51          &       17.45 &  2.31                   & 2.03                     & 4.18                               & DA+M      &      Sec. 4.3.2\\       
     254092090595748096  &     217805062 &             0.90  &        70.84  &        46.70 &       86980$\pm$2390       &       7.23$\pm$0.08 &       12.61 &  3.80                   & 3.93                     & 6.90                               & DAO+BP    &      Sec. 4.3.2\\       
     3875365174618907264 &     122716187 &             1.00  &        154.51 &        7.36  &       27375$\pm$124        &       7.73$\pm$0.02 &       16.60 &  1.20                   & 2.87                     & 4.75                               & DAZ       &      Sec. 4.3.3\\    
     3683519503881169920 &     740711091 &             0.99  &        191.12 &        -1.32 &       22816$\pm$197        &       7.25$\pm$0.01 &       13.99 &  3.62                   & 6.29                     & 8.31                               & DA        &      Sec. 4.3.4\\       
     3800902265750001664 &     400401022 &             0.99  &        174.30 &        3.72  &       6837$\pm$44          &       8.11$\pm$0.02 &       17.72 &  1.35                   & 4.11                     & 3.72                               & DZ        &      Sec. 4.3.4\\       
     3940955205038857728 &     894314169 &             1.00  &        193.10 &        17.95 &       19809$\pm$264        &       7.88$\pm$0.01 &       15.46 &  3.99                   & 2.76                     & 6.62                               & DA        &      Sec. 4.3.4\\       
     4017215231301573376 &     724404116 &             0.99  &        174.47 &        24.86 &       7677$\pm$92          &       7.78$\pm$0.04 &       17.85 &  2.86                   & 4.04                     & 6.65                               & DA        &      Sec. 4.3.4\\       
     634228028075430016  &     583907250 &             0.99  &        141.74 &        19.61 &       8415$\pm$76          &       8.07$\pm$0.12 &       18.58 &  1.38                   & 3.93                     & 6.01                               & DA        &      Sec. 4.3.4\\       
     6963383233077632    &     367309174 &             1.00  &        45.96  &        6.13  &       15150.28$\pm$5399.75 &       8.73$\pm$0.51 &       14.96 &  4.48                   & 13.24                    & 14.71                              & DX        &      Sec. 4.3.4\\       
     800768958390593152  &     299608217 &             1.00  &        145.52 &        38.96 &       15278                &       8.03          &       19.32 &  0.55                   & 4.20                     & 5.95                               & DA        &      Sec. 4.3.4\\       
    \hline
    \end{tabular}
    \addtolength{\tabcolsep}{1pt}
\end{table*}

\begin{figure*} 
    \begin{center}
        \includegraphics[width=0.82\textwidth]{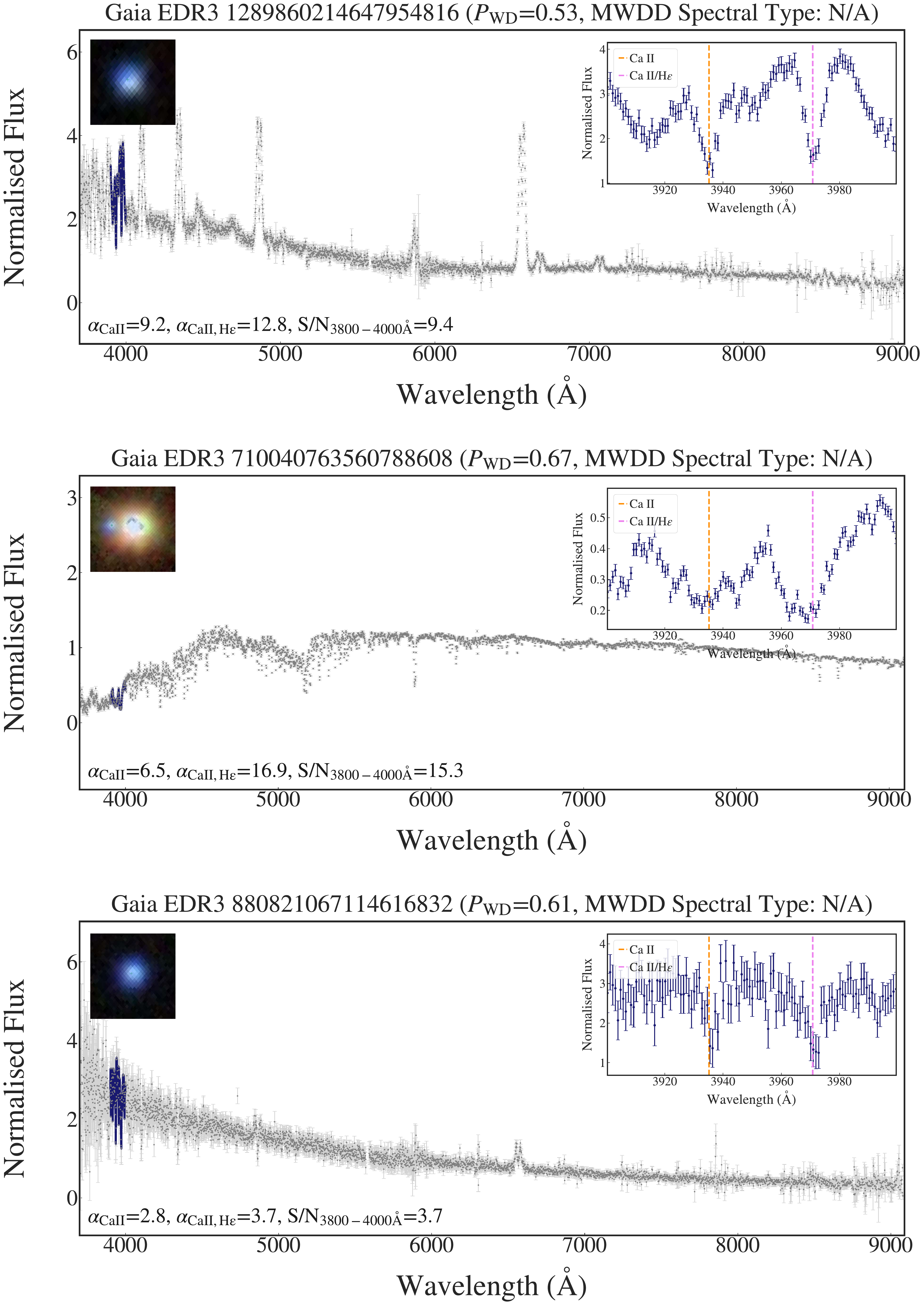}
        \caption{LAMOST low-resolution spectra of \Gaia{} EDR3 1289860214647954816 (top), \Gaia{} EDR3 710040763560788608 (middle), and \Gaia{} EDR3 880821067114616832 (bottom). These three objects were discarded from our final catalogue of polluted white dwarfs after visual inspection of their LAMOST spectra (see Section \ref{sec:visual_check}). Both \Gaia{} EDR3 1289860214647954816 and \Gaia{} EDR3 880821067114616832 exhibit emission features typical of catacylismic variables \citep{Drake:2014, Denisenko:2012}, while \Gaia{} EDR3 710040763560788608 is a blend of a white dwarf and an M-dwarf. In each panel, the inset plots show SDSS DR9 sky images of the stars' field-of-view (top left), as well as a zoomed-in region of their spectra between 3,900-4,000~\angs{}. For each object, we also list their calcium significances and the S/N of their spectrum between 3,800-4,000~\angs{} (see Section \ref{sec:lorentz_classification}).}
        \label{fig:discard_visual}
    \end{center}
\end{figure*}

\begin{figure*} 
    \begin{center}
        \includegraphics[width=0.84\textwidth]{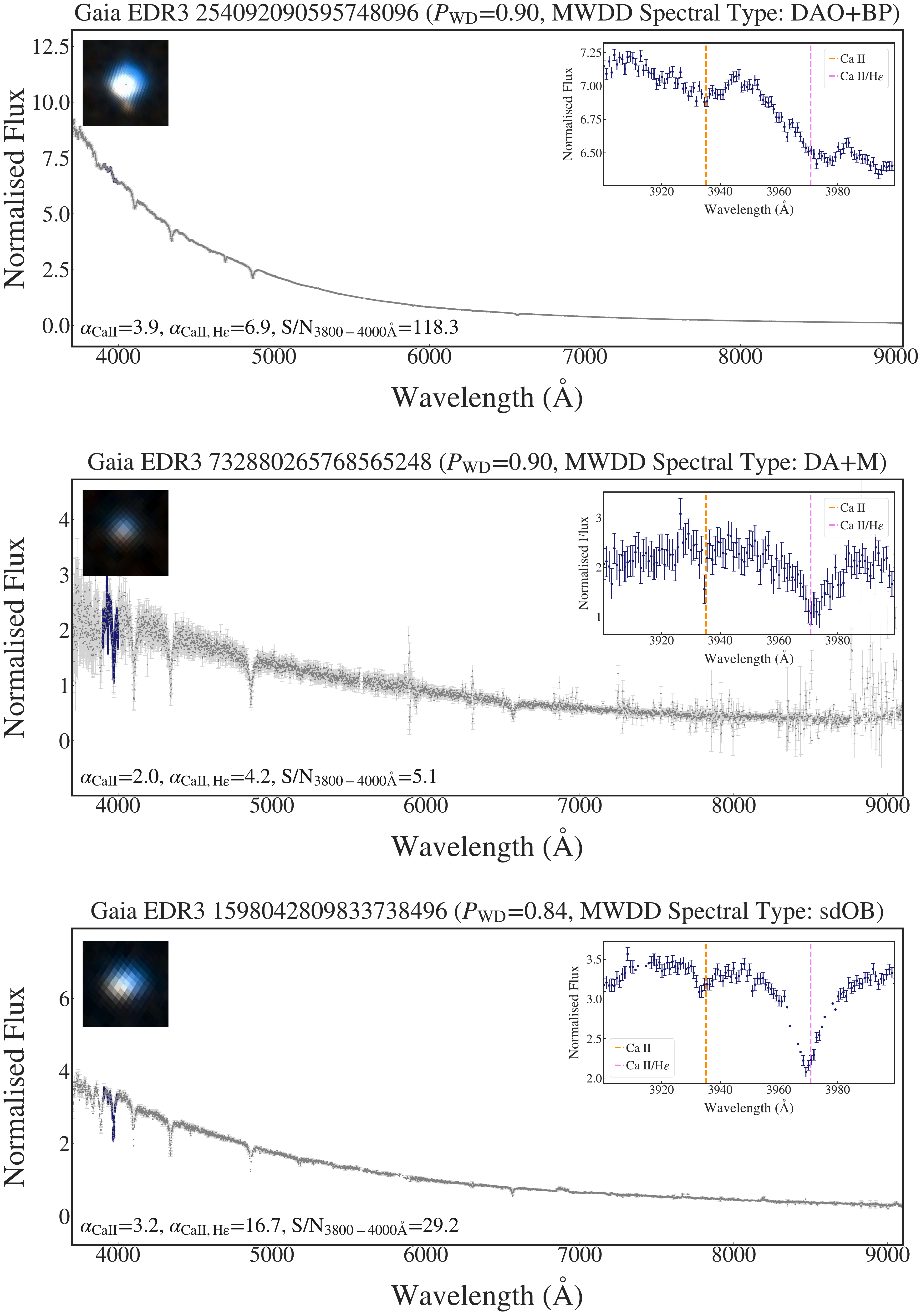}
        \caption{LAMOST low-resolution spectra of \Gaia{} EDR3 254092090595748096 (top), \Gaia{} EDR3 732880265768565248 (middle), and \Gaia{} EDR3 1598042809833738496 (bottom). These three objects were excluded from the final catalogue of polluted white dwarfs during a literature cross-check of their spectral types (see Section \ref{sec:spectraltype_check}). Both \Gaia{} EDR3 254092090595748096 and \Gaia{} EDR3 732880265768565248 are binary systems, while \Gaia{} EDR3 1598042809833738496 is a subdwarf.}
        \label{fig:discard_spectype}
    \end{center}
\end{figure*}

\begin{figure*} 
    \begin{center}
        \includegraphics[width=0.84\textwidth]{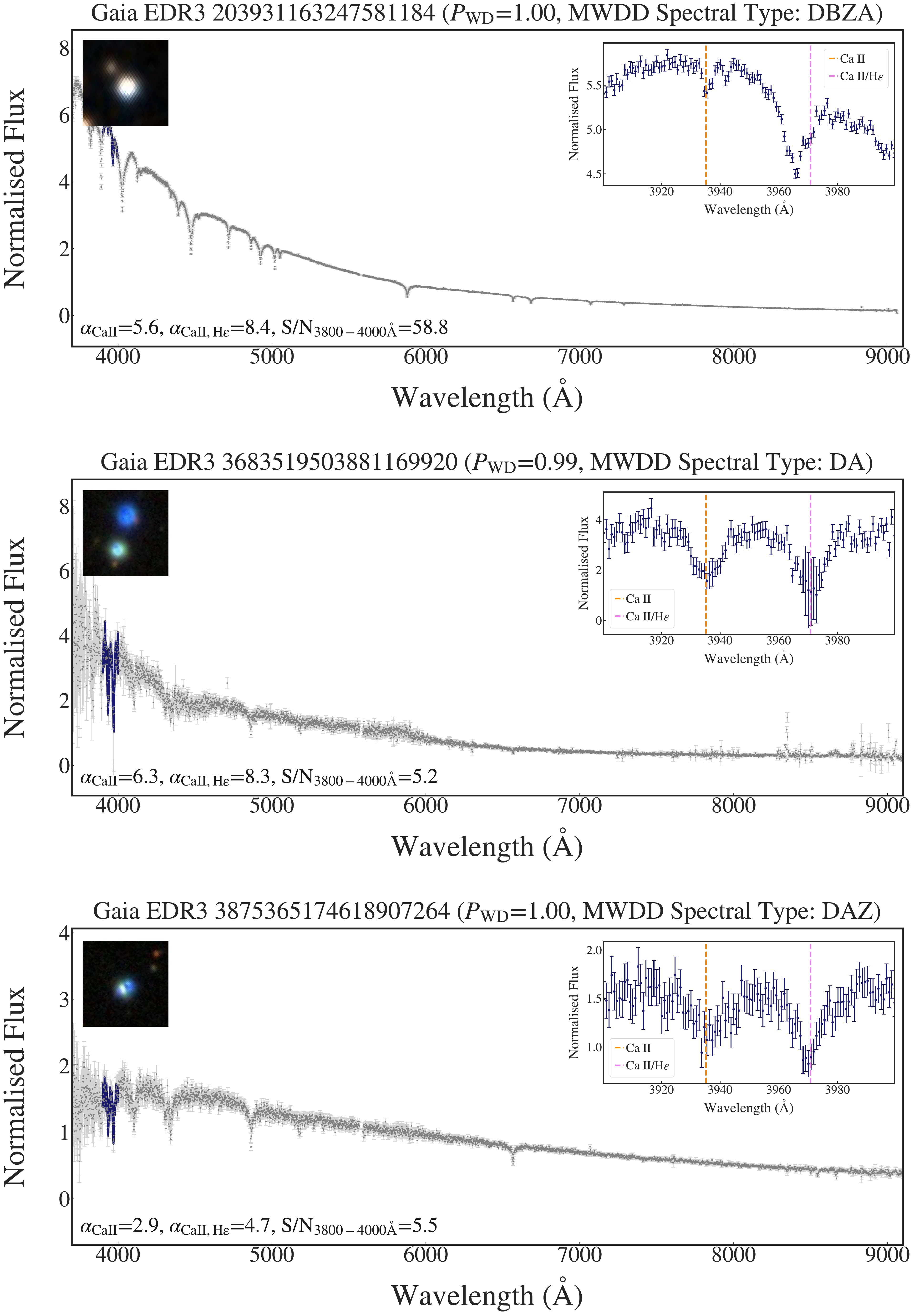}
        \caption{LAMOST low-resolution spectra of \Gaia{} EDR3 203931163247581184 (top), \Gaia{} EDR3 3683519503881169920 (middle), and \Gaia{} EDR3 3875365174618907264 (bottom). As explained in Section \ref{sec:contamination_check}, these stars share their LAMOST field-of-view with another object located, respectively, at a distance of 1.24\arcsec, 3.10\arcsec, and 2.25\arcsec{}. Given that the contaminant of \Gaia{} EDR3 3875365174618907264 is of comparable brightness to the white dwarf ($G_{\rm mag}$=16.6, $G_{\rm mag, comp}$=15.6), we decided to exclude the latter from our final poluted sample. However, we chose to keep \Gaia{} EDR3 203931163247581184 ($G_{\rm mag}$=14.8, $G_{\rm mag, comp}$=21.0) and \Gaia{} EDR3 3683519503881169920 ($G_{\rm mag}$=14.0, $G_{\rm mag, comp}$=20.1), as they are both close to a significantly fainter companion.}
        \label{fig:discard_fibercontamination}
    \end{center}
\end{figure*}

\begin{table*}
    \centering
    \caption{Our full catalogue of GF21/LAMOST white dwarf candidates contains 228 columns from four different sources: the \Gaia{} EDR3 database as reported in GF21, the LAMOST DR9v1 archive, the MWDD, and this work. Here, we only describe the LAMOST columns (Ref. [1]) as well as those added in this paper (Ref. [2]). \label{tab:catalog_format}} 
    \begin{tabular}{llcl}
        Column No.$^{**}$ & Heading & Ref. & Description \\
        \hline 
        \vdots  &   \vdots               & \vdots &    \vdots       \\
        95  & \textsc{ra\_obs}           & [1]    &  Fiber pointing R. A. coordinates of the LAMOST target during its observation (J2000).  \\ 
        96  & \textsc{dec\_obs}          & [1]    &  Fiber pointing Dec. coordinates of the LAMOST target during its observation (J2000).  \\
        97  & \textsc{ra}                & [1]    &  Equatorial R. A. coordinates of the LAMOST target from the LAMOST \\
            &                            &        &  Input Catalogue (in degrees; J2000). \\
        98  & \textsc{dec}               & [1]    &  Equatorial Dec. coordinates of the LAMOST target from the LAMOST \\
            &                            &        &  Input Catalogue (in degrees; J2000). \\
        99  & \textsc{obsid}             & [1]    &  Unique observation ID of a LAMOST spectrum. \\   
        100 & \textsc{uid}               & [1]    &  Unique ID of the LAMOST target based on its equatorial coordinates (in degrees). \\
            &                            &        & from \panstarrs, \Gaia, or LAMOST. \\
        101 & \textsc{gp\_id}            & [1]    &  Unique ID of the LAMOST target based on the equatorial coordinates listed in \textsc{uid}.  \\
        102 & \textsc{designation}       & [1]    &  ID of the LAMOST target in the form JHHMMSS.ss+DDMMSS.ss”, where “HHMMSS.ss” is \\   
            &                            &        &  R.A. (in hours:minutes:seconds) and “+DDMMSS.ss” is Dec. (in degrees:minutes:seconds).  \\ 
        103 & \textsc{obsdate}           & [1]    &  Observation date of the LAMOST target. \\
        104 & \textsc{lmjd}              & [1]    &  Local modified Julian day of the LAMOST observation. \\ 
        105 & \textsc{mjd}               & [1]    &  Modified Julian day of the LAMOST observation. \\ 
        106 & \textsc{planid}            & [1]    &  Plan name of the LAMOST target in use.   \\
        107 & \textsc{spid}              & [1]    &  Positive integer value between 1 and 16 indicating the LAMOST spectrograph ID.  \\
        108 & \textsc{fiberid}           & [1]    &  Positive integer value between 1 and 250 indicating the LAMOST fiber ID.  \\
        109 & \textsc{snru}              & [1]    &  S/N of the LAMOST spectrum in the \textit{u}-band.   \\
        110 & \textsc{snrg}              & [1]    &  S/N of the LAMOST spectrum in the \textit{g}-band. \\
        111 & \textsc{snrr}              & [1]    &  S/N of the LAMOST spectrum in the \textit{r}-band. \\
        112 & \textsc{snri}              & [1]    &  S/N of the LAMOST spectrum in the \textit{i}-band. \\ 
        113 & \textsc{snrz}              & [1]    &  S/N of the LAMOST spectrum in the \textit{z}-band. \\
        114 & \textsc{class}             & [1]    &  Classification of LAMOST target by the LAMOST 1D pipeline. \\
        115 & \textsc{subclass}          & [1]    &  Spectral type of the LAMOST target estimated by the LAMOST 1D pipeline. \\
        116 & \textsc{z}                 & [1]    &  Redshift of the LAMOST target calculated by the LAMOST 1D Pipeline (-9999 if unavailable). \\
        117 & \textsc{z\_err}            & [1]    &  Redshift uncertainty of the LAMOST target calculated by \\
            &                            &        & the LAMOST 1D pipeline (-9999 if unavailable). \\
        118 & \textsc{magtype}           & [1]    &  String field with the available magnitudes of the LAMOST target. \\  
        119 & \textsc{mag1}              & [1]    &  LAMOST \textit{u}-band magnitude.   \\
        120 & \textsc{mag2}              & [1]    &  LAMOST \textit{g}-band magnitude.   \\
        121 & \textsc{mag3}              & [1]    &  LAMOST \textit{r}-band magnitude.    \\
        122 & \textsc{mag4}              & [1]    &  LAMOST \textit{i}-band magnitude. \\   
        123 & \textsc{mag5}              & [1]    &  LAMOST \textit{z}-band magnitude. \\    
        124 & \textsc{mag6}              & [1]    &  LAMOST \textit{j}-band magnitude.  \\
        125 & \textsc{mag7}              & [1]    &  LAMOST \textit{h}-band magnitude.   \\
        126 & \textsc{ps\_id}            & [1]    &  Name of the LAMOST target in the \panstarrs{} database.  \\   
        127 & \textsc{ps\_g}             & [1]    &  \panstarrs{} \textit{g}-band magnitude stored in the LAMOST database. \\
        128 & \textsc{ps\_r}             & [1]    &  \panstarrs{} \textit{r}-band magnitude stored in the LAMOST database.   \\
        129 & \textsc{ps\_i}             & [1]    &  \panstarrs{} \textit{i}-band magnitude stored in the LAMOST database.   \\
        130 & \textsc{ps\_z}             & [1]    &  \panstarrs{} \textit{z}-band magnitude stored in the LAMOST database.  \\
        131 & \textsc{ps\_y}             & [1]    &  \panstarrs{} \textit{y}-band magnitude stored in the LAMOST database.  \\
        132 & \textsc{n\_ps}             & [1]    &  Number of \panstarrs{} objects within  3\arcsec of the LAMOST target.   \\
        133 & \textsc{gaia\_source\_id}  & [1]    &  \Gaia{} DR2 name of the LAMOST target stored in the LAMOST database.  \\   
        134 & \textsc{gaia\_g\_mean\_mag}  & [1]  &  \Gaia{} DR2 \textsc{phot\_g\_mean\_magnitude} value of the \\
            &                              &      &  LAMOST target stored in the LAMOST database. \\
        135 & \textsc{tsource}           & [1]    &  Person or organisation that submitted the LAMOST target to the Input Catalogue. \\
        136 & \textsc{fibertype}         & [1]    &  Fiber type (Obj, Sky, F-std, Unused, PosErr, Dead).$^{\mathsection}$ \\ 
        137 & \textsc{tfrom}             & [1]    &  LAMOST Input Catalogue submitted by the person/organisation, as described by \textsc{tsource}. \\  
        138 & \textsc{tcomment}          & [1]    &   Target ID from an external catalogue (e.g. SDSS, UCAC4, \panstarrs).\\
        139 & \textsc{offsets}           & [1]    & Boolean value indicating if there was a fiber offset during the LAMOST observation. \\   
        140 & \textsc{offsets\_v}        & [1]    &  If \textsc{offsets} is True, offset distance (in arcseconds) from the target's\\
            &                            &        & coordinates in the LAMOST Input Catalogue. \\
        141 & \textsc{fibermask}         & [1]    & Integer value indicating possible fiber problems; if 0, the fiber worked well. \\ 
        142 & \textsc{with\_norm\_flux}  & [1]    &  Boolean value. If 1, the FITS file of the LAMOST target contains its continuum-normalised \\
            &                            &        &  spectrum; if 0, the flux is not continuum-normalised.   \\
        190 & \textsc{GaiaLAMOST\_Separation}   & [2] &   Separation distance (in arcseconds) between the LAMOST target and the nearest \\
            &                                   &     &  \Gaia{} EDR3 white dwarf candidate in the catalogue of GF21. \\
        \vdots   &        \vdots                     & \vdots  &       \vdots       \\
        \hline
    \end{tabular}
\end{table*}

\begin{table*}
    \centering
	\contcaption{Table \ref{tab:catalog_format}. }
	\begin{tabular}{llcl}
        Column No.$^{\dagger}$ & Heading & Ref. & Description \\
        \hline 
          \vdots   &        \vdots                     & \vdots  &       \vdots       \\
        191 & \textsc{GroupID}                  & [2] &   Integer value from our \texttt{TOPCAT} cross-matching routine identifying  all the \\
            &                                   &     & LAMOST objects associated to a given GF21 star (Section \ref{sec:methodology_creation_catalog}). \\ 
            &                                   &     & If NaN, the GF21/LAMOST match is unique. \\
        192 & \textsc{GroupSize}                & [2] &  Integer value from our \texttt{TOPCAT} cross-matching routine indicating the \\
            &                                   &     &  number of stars in each \textsc{GroupID} (Section \ref{sec:methodology_creation_catalog}). \\
        193 & \textsc{lamost\_snr\_mean}        & [2] &  Mean S/N of the LAMOST spectrum; (Section \ref{sec:properties_catalogue}).    \\ 
        194 & \textsc{snr\_ca\_window}          & [2] &  S/N of the wavelength window between 3,800~\angs{} and 4,000~\angs{} (Section \ref{sec:lorentz_classification}). \\
        195 & \textsc{sig\_H}                   & [2] &  Statistical significance of the H8 line at $\sim$3,890~\angs{} (Section \ref{sec:lorentz_classification}).   \\
        196 & \textsc{sig\_Ca}                  & [2] &  Statistical significance of the Ca II line at $\sim$3,934~\angs{} (Section \ref{sec:lorentz_classification}).   \\
        197 & \textsc{sig\_CaH}                 & [2] &  Statistical significance of the combined Ca II and H$\epsilon$ lines at $\sim$3,969.5~\angs{} (Section \ref{sec:lorentz_classification}). \\
        198 & \textsc{CaPollution}              & [2] &  Presence of calcium photospheric pollution between 3,900~\angs{} and 4,000~\angs{} (Section \ref{sec:final_polluted_sample}). \\
        199 & \textsc{SpectralClass}            & [2] &  Spectral classification of the star based on the strength of the observed. \\
             &                                  &     &  absorption lines in its LAMOST spectrum. \\
        200 & \textsc{mg\_detection}            & [2] &  Visual assessment of the presence of magnesium pollution (``Y'', ``N'', ``A'' for Yes, No, \\
             &                                  &     &  and Ambiguous; see Section \ref{sec:other_elements}).  \\
        201 & \textsc{contamination\_1.5arcsec} & [2] &  Possible contamination from \Gaia{} objects located within 1.5\arcsec{} of  \\
            &                                   &     &  the LAMOST target (``Y'' or ``N/A''; see Section \ref{sec:methodology_creation_catalog}).   \\
        202 & \textsc{Num\_HST\_Data}           & [2] &  Number of existing HST observations acquired with the STIS/CCD instrument.  \\
        203 & \textsc{TIC\_distance\_arcsec}    & [2] &  Separation distance between the LAMOST target and the\\
            &                                   &     &  nearest TESS source (in arcseconds). \\
        204 & \textsc{TIC\_Vmag}                & [2] &  TIC \textit{V}-band magnitude.\\   
        205 & \textsc{TIC\_Vmag\_err}           & [2] &  TIC \textit{V}-band magnitude uncertainty.\\  
        206 & \textsc{TIC\_Bmag}                & [2] &  TIC \textit{B}-band magnitude.    \\
        207 & \textsc{TIC\_Bmag\_err}           & [2] &  TIC \textit{B}-band magnitude uncertainty.\\  
        208 & \textsc{TIC\_Hmag}                & [2] &  TIC \textit{H}-band magnitude.    \\
        209 & \textsc{TIC\_Hmag\_err}           & [2] &  TIC \textit{H}-band magnitude uncertainty.\\  
        210 & \textsc{TIC\_Kmag}                & [2] &  TIC \textit{K}-band magnitude.    \\
        211 & \textsc{TIC\_Kmag\_err}           & [2] &  TIC \textit{K}-band magnitude uncertainty.\\
        212 & \textsc{GALEX\_distance\_arcmin}  & [2] &  Separation distance between the LAMOST target and \\
            &                                   &     & the nearest \GALEX{} source (in arcminutes). \\   
        213 & \textsc{GALEX\_Fuv\_mag}          & [2] &  \GALEX{} FUV magnitude.    \\
        214 & \textsc{GALEX\_Fuv\_magerr}       & [2] &  \GALEX{} FUV magnitude uncertainty.    \\
        215 & \textsc{GALEX\_Nuv\_mag}          & [2] &  \GALEX{} NUV magnitude.    \\
        216 & \textsc{GALEX\_Nuv\_magerr}       & [2] &  \GALEX{} NUV magnitude uncertainty.    \\  
        217 & \textsc{PANSTARRS\_Distance\_deg} & [2] &  Separation distance between the LAMOST target and  \\
            &                                   &     & the nearest \panstarrs{} source (in degrees). \\
        218 & \textsc{PANSTARRS\_gMeanPSFMag}    & [2] & \panstarrs{} mean \textit{g}-magnitude.   \\
        219 & \textsc{PANSTARRS\_gMeanPSFMagErr} & [2] & \panstarrs{} mean \textit{g}-magnitude uncertainty.   \\
        220 & \textsc{PANSTARRS\_rMeanPSFMag}    & [2] & \panstarrs{} mean \textit{r}-magnitude.   \\
        221 & \textsc{PANSTARRS\_rMeanPSFMagErr} & [2] & \panstarrs{} mean \textit{r}-magnitude uncertainty. \\
        222 & \textsc{PANSTARRS\_iMeanPSFMag}    & [2] & \panstarrs{} mean \textit{i}-magnitude.   \\
        223 & \textsc{PANSTARRS\_iMeanPSFMagErr} & [2] & \panstarrs{} mean \textit{i}-magnitude uncertainty.  \\ 
        224 & \textsc{PANSTARRS\_zMeanPSFMag}    & [2] & \panstarrs{} mean \textit{z}-magnitude.   \\
        225 & \textsc{PANSTARRS\_zMeanPSFMagErr} & [2] & \panstarrs{} mean \textit{z}-magnitude uncertainty.       \\
        226 & \textsc{PANSTARRS\_yMeanPSFMag}    & [2] & \panstarrs{} mean \textit{y}-magnitude.   \\
        227 & \textsc{PANSTARRS\_yMeanPSFMagErr} & [2] & \panstarrs{} mean \textit{y}-magnitude uncertainty.          \\
        \hline
    \end{tabular}
    \vspace{-1.2pt}
    \begin{quote}
      \hspace{0.4cm} [$^{**}$]: We use the Pythonic convention, which assumes that the first column of our catalogue is in index 0. \\
      \hspace{0.52cm} [$^{\mathsection}$\textit{Obj}]: The fiber is assigned to an astrophysical object (e.g. stars, galaxies). \textit{Sky}: The fiber is used to take sky flats. \text{F-std}: The fiber is used to take flux standards of a calibration star. \textit{Unused}: Unused fiber. \textit{PosErr}: Wrong fiber position. \textit{Dead}: Fiber no longer in operation.
    \end{quote}
\end{table*}

\begin{table*}
    \small
    \caption{\Gaia{} EDR3 white dwarf candidates discarded in our atmospheric analysis (see Section \ref{sec:atmospheric_analysis}). The superscript letters indicate the discovery paper of the source.}
    \begin{tabular}{lccc} 
        \hline
        \Gaia{} EDR3                        &  P$_{\rm WD}$ & Issue \\
        \hline
        \hline
        340886198462842112$^{\tiny{a}}$      & 0.99  &   \Teff{}$\geq$25,000~K based on preliminary fits to their existing \panstarrs{} or \Gaia{} photometry.\\
        480570075502703488$^{\tiny{b}}$      & 0.75  &   \Teff{}$\geq$25,000~K based on preliminary fits to their existing \panstarrs{} or \Gaia{} photometry. \\
        27578539058961280$^{\tiny{c}}$       & 1.00  & No converged spectroscopic solution (potential unresolved binary?)  \\
        295237976073772672$^{\tiny{b}}$      & 0.95  &  No converged spectroscopic solution (potential unresolved binary?)\\
        917141857485288704$^{\tiny{c}}$      & 0.93  &  No converged spectroscopic solution (potential unresolved binary?) \\
        \hline          
    \end{tabular}
     \begin{quote}
      \hspace{0.4cm} White Dwarf Discovery Reference: [a]: \citealt{GentileFusillo:2019}, [b]: \citealt{GentileFusillo:2021}, [c]: \citealt{Jimenez:2018}.
    \end{quote}
    \label{tab:discarded_in_ca_fits}
\end{table*}


\bsp	
\label{lastpage}
\end{document}